\documentclass[lettersize,journal]{IEEEtran}
\usepackage{amsmath,amsfonts}
\usepackage{algorithmic}
\usepackage{algorithm}
\usepackage{array}
\usepackage{amssymb}
\usepackage{textcomp}
\usepackage{stfloats}
\usepackage{url}
\usepackage{verbatim}
\usepackage{graphicx}
\usepackage{multirow}
\usepackage{makecell}
\usepackage{diagbox}
\usepackage[labelformat=simple]{subcaption} 
 
\usepackage{caption}
\usepackage{cite}
\usepackage{booktabs}
\usepackage{svg}
\usepackage{titlesec}
\titlespacing{\section}{0pt}{0.5\baselineskip}{0.3\baselineskip}
\titlespacing{\subsection}{0pt}{0.4\baselineskip}{0.2\baselineskip}
\titlespacing{\subsubsection}{0pt}{0.3\baselineskip}{0.1\baselineskip}

\usepackage{etoolbox}

\usepackage{float}
\usepackage{microtype}

\begin{document}

\title{mmAlert: A Simultaneous Device Localization and Target Tracking System via Cooperative Passive Sensing }

\author{
Chao Yu, Bojie Lv, Chunxi Chen, Jingwen Zhang and Rui Wang
\thanks{Chao Yu, Bojie Lv and Rui Wang are with the Department of Electrical and Electronic Engineering, Southern University of Science and Technology, Shenzhen, China. Chunxi Chen and Jingwen Zhang are with the College of Semiconductors (National Graduate College for Engineers), Southern University of Science and Technology, Shenzhen, China.

This work has been accepted for publication in IEEE Transactions on Wireless Communications (TWC). DOI: 10.1109/TWC.2026.3698942}
}

\maketitle

\begin{abstract}
In this paper, a cooperative passive sensing system in millimeter-wave (mmWave) band for simultaneous device localization and target tracking, namely mmAlert, is proposed. Specifically, in uplink communication with at least two transmitters, the receiver receives the line-of-sight (LoS) signals and the scattered signals off a moving target, respectively. Based on the received signals of the sensing time intervals, when a passive target moves along one or multiple unknown trajectories, mmAlert could measure the angles-of-arrival (AoAs) and bistatic Doppler frequencies of the echoes from the sensing target, and then jointly estimate the locations of the transmitters and the trajectories of the target. Specifically, the transmitters' locations and the moving target's trajectories can be searched by minimizing the weighted mean squared error of the AoA and Doppler measurements. The optimal solution of the minimization problem is prohibitive due to the large number of variables. Hence, a low-complexity algorithm based on the alternating optimization is proposed, where the extended Kalman filter (EKF) is introduced to quickly shape the trajectories. The mmAlert is implemented in a 60GHz communication testbed. The experiment shows with the received signal spanning a single trajectory, the average localization error of the transmitters and average trajectory reconstruction error are 0.76 m and 0.29 m, respectively. The average errors are suppressed to 0.07 m and 0.2 m respectively, if the received signal spanning 50 trajectories is used. This justifies the benefit of trajectory diversity in localization and tracking.
\end{abstract}


\section{Introduction}
\IEEEPARstart{M}{mWave} communication is a key technology for future wireless communication systems due to its abundant spectrum resource \cite{sec1_1,mmwave_2,mmwave}. However, the short wavelength of mmWave signals also brings inherent disadvantages, such as severe signal attenuation, susceptibility to obstacles, and frequent beam alignment \cite{mmwave,sec1_2,mmWave_beam}. Fortunately, the integrated sensing and communication (ISAC) technology offers a promising solution to relieve the above issues \cite{sec1_4,sec1_5,mmWave_ISAC}. For instance, it is shown that the wireless communication systems could locate moving obstacles \cite{wifi_tracking1,wifi_tracking2} or its devices \cite{device_loc1,device_loc2} via sensing techniques. 
Hence, potential mmWave signal blockage can be predicted, and fast beam alignment can be facilitated via location information. 
In the existing literature, device localization and passive target (i.e., a target that does not emit signals) tracking are investigated in a separate manner. The prerequisites of the existing techniques might be stringent. For example, tracking of passive target requires prior knowledge on the locations of multiple coordinated sensing devices, whereas localization of the device itself relies on either collaborative positioning using multiple receivers or wideband signals for accurate ranging.
In this paper, we would like to show that the tracking of passive moving target and the localization of communication devices might facilitate each other in an mmWave communication system, when either task cannot be accomplished solely. 

\subsection{Device Localization}
In the existing literature, the localization of communication devices with inband sensing usually relies on the measurement of signal fingerprints in advance \cite{RSSI, CSI, wifi_fingerprint} or prior knowledge on the locations of multiple anchor devices (e.g., access points or base stations) \cite{AoA,AoA_2,ToF_2,RTT}.

For example, in \cite{RSSI}, a fingerprint database of received signal strength indicator (RSSI) was constructed for device localization, where a localization and tracking algorithm based on recurrent neural networks (RNNs) was proposed. It leveraged the continuity of device movement to improve localization accuracy.
Moreover, since the channel state information (CSI) contains more fine-grained environmental information, the fingerprint based on CSI is expected to yield higher localization accuracy \cite{CSI, wifi_fingerprint}. For example, in \cite{wifi_fingerprint}, a novel indoor wireless localization algorithm based on a broad learning system (BLS) was proposed, which achieved meter-level localization accuracy using CSI fingerprint.
However, all the fingerprint-based methods require a significant effort of collecting signal features in advance. Moreover, the changes in the environment, e.g., walking persons, might severely degrade the localization accuracy.

Wireless communication devices can also be localized based on measurements of angle-of-arrival (AoA) \cite{AoA,AoA_2} and time-of-flight (ToF) \cite{ToF_2,RTT,Multi_carrier} at multiple anchor devices, whose locations are known in advance.
For example, in \cite{AoA}, it was proposed to first estimate the AoAs from three multi-antenna access points (APs) and then infer the target position using triangulation. The algorithm achieved a median positioning error of 57 cm. In \cite{AoA_2}, a joint AoA estimation method with multiple APs was proposed. Considering the coupling between the AoA measurements of multiple APs with respect to the same target, the AoAs should not be estimated independently. Instead, AoA estimation at all the APs was modeled as a joint parametric optimization problem to improve the localization accuracy. The ToF of a communication signal is usually measured by the round trip time (RTT). A trilateration algorithm based on the RTT was proposed in \cite{RTT} for device localization. It was shown that the average localization error was 1.2 m when the Wi-Fi signal bandwidth was 40MHz. 
In addition to the RTT, the ToF can also be measured via the phase difference of a multi-carrier system. For example, a multi-carrier ranging algorithm based on orthogonal frequency division multiplexing (OFDM) signals was proposed in \cite{Multi_carrier}. It was shown that the median ranging errors of 0.65 m and 0.98 m could be achieved in line-of-sight (LoS) and non-line-of-sight (NLoS) scenarios, respectively.
Nevertheless, it relied on a complicated calibration algorithm to mitigate the effect of carrier frequency offset (CFO) between the transmitter and receiver. In all the above works, prior location knowledge on multiple anchor devices is necessary. Moreover, stringent requirements, such as large signal bandwidth and fine synchronization among the devices, should be satisfied in the measurement of ToF.
In \cite{MonoLoco}, the multipath relative time-of-flight (rToF) was exploited to estimate the triangle between the LoS path and reflected paths, such that the transmitter could be localized. In this work, sufficiently large signal bandwidth (40MHz) was necessary to distinguish reflected paths from the LoS path in the channel impulse response (CIR). Although it did not rely on the prior location knowledge, specific NLoS environment was required.

\subsection{Trajectory Reconstruction}
There have been numerous studies investigating the trajectory reconstruction of passive moving targets via half-duplex communication devices. For example, the Widar system proposed in \cite{widar} tracked human motion trajectories by detecting bistatic Doppler frequencies through CSI. It could achieve a decimeter-level accuracy with at least 6 receive RF chains. The IndoTrack system proposed in \cite{Indotrack} leveraged bistatic Doppler frequency and AoA measurements of 6 receive RF chains for trajectory reconstruction, achieving an error margin within 0.48 m. The WiDFS system proposed in \cite{wifi_tracking1} utilized three antennas on a single receiver to cancel the carrier spectrum interference, such that the bistatic Doppler frequency and AoA information could be estimated for tracking. An average tracking error of 0.72 m was verified by the experimental results. Additionally, the Witraj system proposed in \cite{Witraj} relied solely on bistatic Doppler frequencies in trajectory reconstruction. When the positions of the sensing targets at the very beginning, one transmitter and three receivers were known, Witraj achieved a median trajectory error of 0.3 m. An ML-Track algorithm was proposed in \cite{ML_Track}, which employed a round-trip CSI mechanism to eliminate the impact of CFO in Doppler frequency estimation. The experimental results demonstrated that under 4-link conditions, the algorithm attained a median error of 0.23 m.

Passive sensing is another promising approach for trajectory tracking using half-duplex communication transceivers. For instance, it was demonstrated in \cite{PassiveSensing} that human movements behind walls could be tracked by exploiting Wi-Fi signals in passive sensing. Meanwhile, a passive sensing system based on mmWave communication signals was proposed in \cite{PassiveHand} to achieve millimeter-level handwriting tracking. A passive sensing system based on long-term evolution (LTE) signals was proposed in \cite{UAV}, which achieved a meter-level tracking accuracy for a low-altitude drone flying hundreds of meters away from the base station. In all the above works, the locations of transmitters and receivers are known in advance.

\subsection{Our Contributions}
As discussed in the previous parts, the localization of communication devices and the trajectory tracking of passive targets are treated separately in the existing literature. 
In fact, the measurements of AoA, ToF, and Doppler frequency for trajectory tracking strongly depend on the locations of transmitters and receivers in ISAC systems. This facilitates the joint design of passive target tracking and device localization. Moreover, it is much easier to detect the AoA and Doppler frequency of a wireless communication signal than its propagation distance, due to the limited signal bandwidth and asynchronous clocks between the transmitter and the receiver. Hence, it is interesting to raise the following questions:
\begin{itemize}
    \item Could the device localization and passive target tracking be conducted jointly? Or could the trajectory of a passive target be tracked in an ISAC system without the location information of the communication devices?
    \item Could the above joint estimation be conducted with only angle and Doppler measurements?
\end{itemize}

In this paper, we would like to shed some lights on the above questions by proposing a method of simultaneous localization and tracking (SLAT) for an mmWave ISAC system, namely mmAlert. Specifically, at least two transmitters simultaneously transmit uplink signals to one receiver in different frequency bands, and there is one moving target in the communication environment. The transmitters' locations and the trajectories of the moving target with respect to the receiver are all unknown parameters, which usually happens in communication systems. In this scenario, the existing target tracking methods cannot be applied due to the lack of communication devices' locations; the localization of communication devices suffers from sampling and carrier frequency offsets. Instead, the proposed mmAlert is capable of estimating the above unknown parameters with high accuracy. The estimation is facilitated by the movement of the target. Multiple trajectories of the target are sensed. During the sensing phase, the receiver collects the LoS signal from the two transmitters and the scattered signal off the sensing target, such that the AoA and Doppler frequency of the scattered signal can be detected. Based on these detections of multiple trajectories, the SLAT design can be formulated as an optimization problem: find the best locations of the transmitters and the trajectories of the sensing target to minimize the weighted summation of AoA and Doppler measurement errors. The main contributions of this paper are summarized below.
\begin{enumerate}
\item To the best of our knowledge, this is the first work exploiting the joint detection of communication devices' locations and multiple sensing targets' trajectories. Although the mmAlert is implemented in an mmWave communication system, the principle could be extended to other communication systems, e.g., the Wi-Fi system. Compared with our prior work \cite{MS_LM} on SLAT for a single trajectory, the trajectory diversity is exploited in this work, such that the tracking and localization accuracy is significantly improved. 
\item The above optimization problem for joint detection has not been investigated before, owing to the novelty of the joint design. Due to the high dimensionality of the variable space, obtaining the optimal solution is computationally prohibitive. Therefore, a suboptimal, low-complexity solution framework integrating alternating optimization (AO) and the extended Kalman filter (EKF) is proposed. Particularly, the optimization variables consist of the transmitters' locations, the initial positions and shapes of the trajectories. Given the locations, the initial position and shape of each trajectory could be optimized, respectively; and given multiple trajectories, the transmitters could be located with a good accuracy. In the former step, given the initial position of a trajectory, the trajectory shape could be obtained via EKF; and given the trajectory shape, the gradient of the initial position could be derived analytically. 
\item The performance of the proposed SLAT design is demonstrated in a 60GHz ISAC testbed, whose measurements are released for open access\footnote{https://github.com/lasso-sustech/mmAlert}.
\end{enumerate}
Experimental results demonstrate that mmAlert achieves an average trajectory reconstruction error of 0.29 m only using the measurements of a single trajectory. When the measurements on multiple trajectories are available, the transmitter localization error reduces from 0.76 m to 0.07 m, and the trajectory reconstruction error decreases from 0.29 m to 0.2 m.

The remainder of this paper is structured as follows. In Section \ref{Sec:system_overview}, an overview of the system is provided. In Section \ref{sec:system_model}, the system's motion model and uplink transmission model are elaborated. In Section \ref{sec:Doppler-AoA}, the estimation methods for bistatic Doppler frequency and AoA are described. In Section \ref{sec:Problem Formulation}, the joint estimation problem in SLAT design is formulated, then in Section \ref{sec:method}, a low-complexity solution algorithm is proposed. In Section \ref{sec:Experiment}, the experimental results and analysis are presented. Finally, the conclusion is drawn in Section \ref{sec:conclusions}.

\section{Overview of mmAlert System} \label{Sec:system_overview}
\begin{figure}[htbp] 
	\centering  
	\includegraphics[width=0.95\columnwidth]{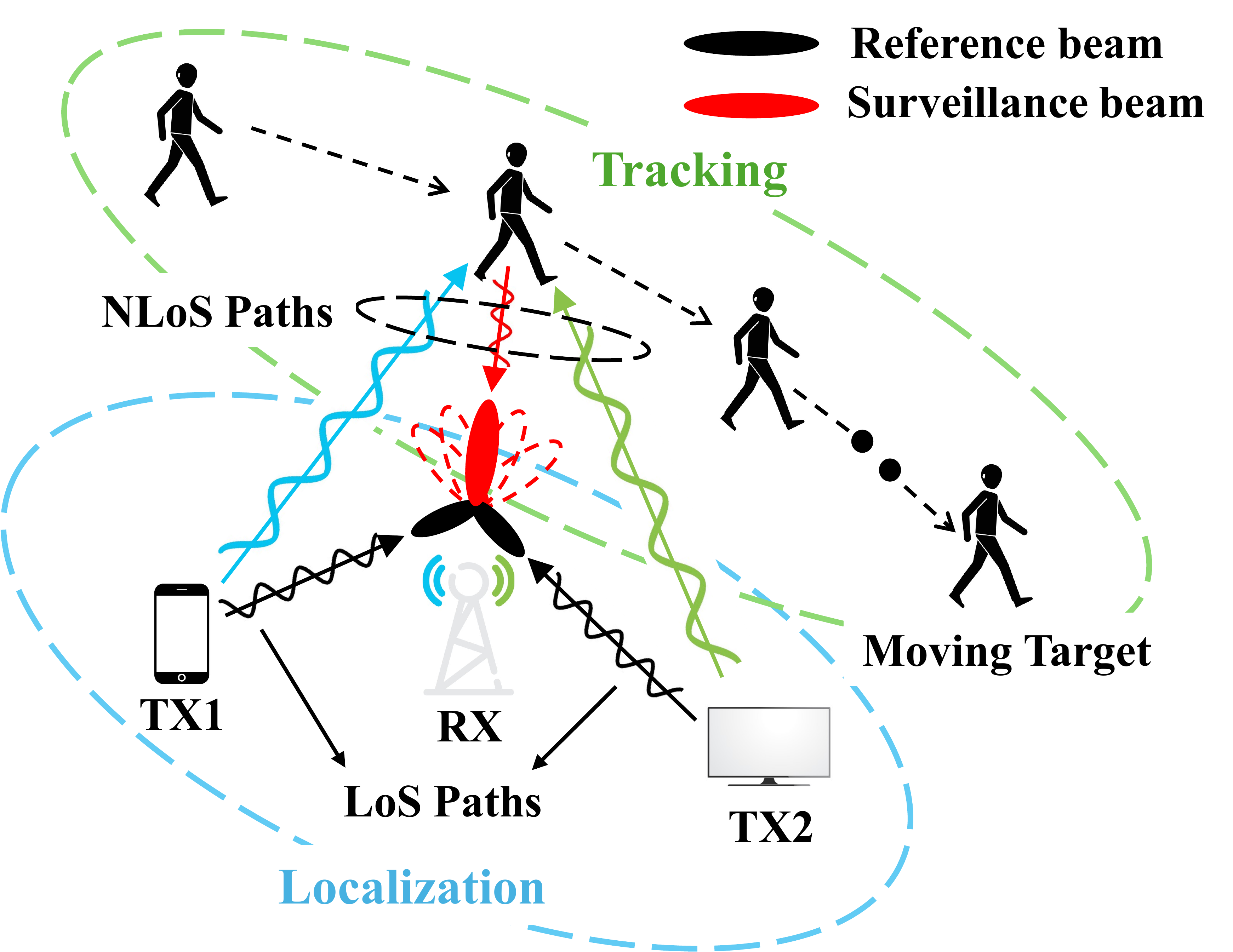}
	\caption{Illustration of deployment scenario of the mmAlert system.}
	\label{fig: 01_overview}
\end{figure}
In this paper, a simultaneous localization and tracking (SLAT) system in mmWave band, namely mmAlert, is proposed. The mmAlert system can be deployed in the mmWave uplink communication scenario with at least two transmitters (user equipments), one receiver (base station) and a moving target, as illustrated in Fig. \ref{fig: 01_overview}. It enables the localization of the transmitters and the trajectory reconstruction of the moving target via communication signals. Such that potential link blockage can be predicted. With the localization and tracking results, the human motion recognition with the mmWave communication signal could also be facilitated. 

In mmAlert, one phased array is adopted at each transmitter for transmit beamforming, and three phased arrays \footnote{In fact, the three phased arrays can be replaced by one hybrid antenna array as in \cite{hybrid_beam}, which provides multiple RF chains and multiple beams in one single antenna array.} are adopted at the receiver for joint sensing and communication. The receiver receives signals from the two transmitters in two orthogonal frequency bands, respectively. In data transmission, the LoS path \footnote{It should be noted that mmAlert does not require that the reference channel follows a LoS path, a strong and static NLoS path can also works.} from each transmitter to the receiver is dominant. Hence, the receiver aligns two of its receive beams with the LoS paths from the two transmitters, respectively. Meanwhile, the uplink signal from each transmitter may scatter at the moving target, carrying the motion information of the target. The third receive beam of the receiver rotates periodically to capture the scattered signal from the moving target in both frequency bands, yielding a NLoS propagation path. The signal-to-noise ratio (SNR) of the NLoS path is significantly weaker than that of the LoS path, especially in mmWave band. Hence, there is no benefit of exploiting the NLoS signal in data communication. However, the proposed mmAlert could utilize the weak NLoS signal to locate the transmitters and reconstruct the trajectory of the moving target. 

From the perspective of passive sensing, LoS paths and NLoS paths scattered off the moving target are referred to as the reference and surveillance channels, respectively, and the corresponding receive beams are referred to as the reference and surveillance beams respectively. Denote the two transmitters as transmitter 1 and 2, the two reference beams as reference beam 1 and 2, the reference and surveillance channels from the transmitter 1 and 2 as the reference channel 1 and 2, surveillance channel 1 and 2, respectively. On one hand, the bistatic Doppler frequencies (the Doppler frequencies of the surveillance channels) can be estimated by comparing the receive signals of the surveillance and the reference channels. As a remark, due to the carrier frequency offset (CFO) between the transmitters and the receiver, the bistatic Doppler frequencies could hardly be accurately measured solely via the surveillance channels. The reference channel could help to eliminate the effect of CFO \cite{PassiveSensing, PassiveHand}. On the other hand, since the surveillance beam rotates periodically, its direction with the maximum echo signal power can be treated as the AoA of both surveillance channels (Note that the AoAs of both surveillance channels are identical).

Without prior knowledge of the target's initial location or the transmitters' positions, mmAlert utilizes only short term observations of the bistatic Doppler frequencies and AoAs from two surveillance channels. It employs a low‑complexity search algorithm to jointly estimate the positions of both transmitters and the trajectory of the moving target relative to the receiver, achieving a close match to the observations with minor error.

Moreover, if multiple trajectories of the moving target could be collected in different time periods \footnote{For example, different persons walk through the communication environment in different time periods.}, the diversity of these trajectories can be leveraged. Under the condition that the locations of the two transmitters are fixed, such diversity helps improve both the localization accuracy of the transmitters and the reconstruction accuracy of the trajectories.
Finally, as a remark, although the mmAlert is designed for FDMA (or OFDMA) uplink systems, the method can also be extended to TDMA or CDMA systems, as long as the signals of the two reference channels and the two surveillance channels can be separated.

\section{System Model} \label{sec:system_model}

\begin{figure}[htbp]
	\centering
	\includegraphics[width=0.95\columnwidth]{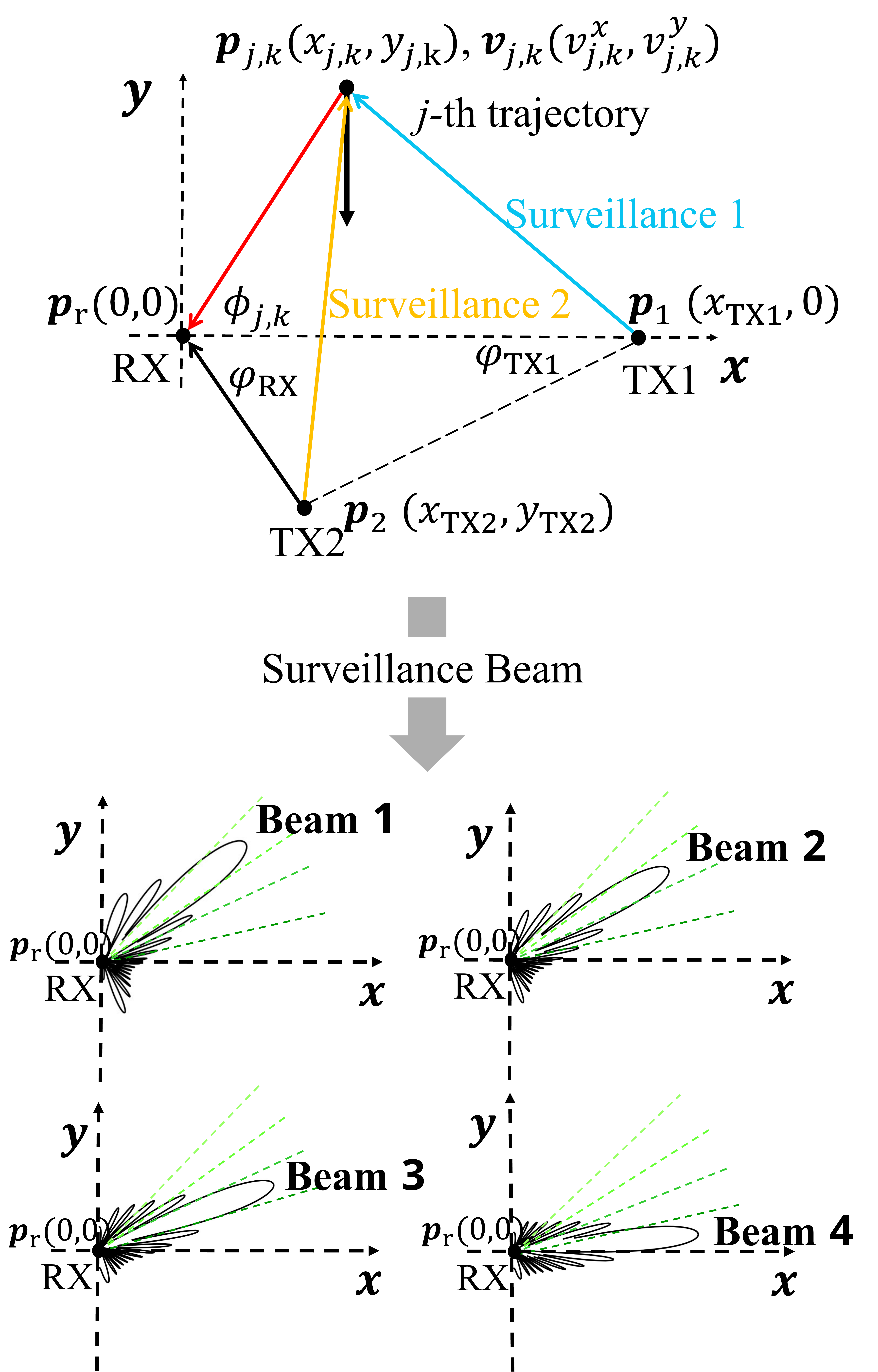}
	\caption{Illustration of SLAT scenario with the number of surveillance beam directions Q=4.}
	\label{fig:02_coordinate system}
\end{figure}

Without loss of generality, the positions of the transmitter 1, 2 and the receiver are denoted as $\boldsymbol{p}_{\mathrm{1}} = [x_{\mathrm{TX1}},0]^{\mathsf{T}}, \boldsymbol{p}_{\mathrm{2}} = [x_{\mathrm{TX2}},y_{\mathrm{TX2}}]^{\mathsf{T}}$ and $\boldsymbol{p}_{\mathrm{r}} = [0,0]^{\mathsf{T}}$, respectively, as illustrated in Fig. \ref{fig:02_coordinate system}. Based on the existing multiple signal classification (MUSIC) algorithm \cite{multone,JCIN}, which is applicable to analog MIMO architectures, the receiver can estimate the signal AoA from the two transmitters, and align two of its receive beams, namely reference beams 1 and 2, with reference channels 1 and 2, respectively.
Hence, the angle difference of arrival $\varphi_{\mathrm{RX}}$ can be calculated from the two AoAs. Similarly, the angle $\varphi_{\mathrm{TX1}}$ in Fig. \ref{fig:02_coordinate system} can also be estimated by swapping the roles of transmitter and receiver. Hence, it is assumed that the angles of $\varphi_{\mathrm{RX}}$ and $\varphi_{\mathrm{TX1}}$ are known in this paper. As a remark, when the positions of the transmitters change, both $\varphi_{\mathrm{RX}}$ and $\varphi_{\mathrm{TX1}}$ should be recalculated. Therefore, both $\varphi_{\mathrm{RX}}$ and $\varphi_{\mathrm{TX1}}$ are known to the mmAlert system. According to the property of triangle, $\boldsymbol{p}_{\mathrm{2}}$ can be determined by
\begin{align}
    \left \{
    \begin{aligned}
        x_{\mathrm{TX2}} = \dfrac{\sin{\varphi_{\mathrm{TX1}}}\cos{\varphi_{\mathrm{RX}}}}{\sin({\varphi_{\mathrm{TX1}} + \varphi_{\mathrm{RX}}})} x_{\mathrm{TX1}} \\[5pt]
        y_{\mathrm{TX2}} = \dfrac{\sin{\varphi_{\mathrm{TX1}}}\sin{\varphi_{\mathrm{RX}}}}{\sin({\varphi_{\mathrm{TX1}} + \varphi_{\mathrm{RX}}})} x_{\mathrm{TX1}}
    \end{aligned}
    \right . ,\label{eqn:tri}
\end{align}
as long as $x_{\mathrm{TX1}}$ can be estimated. 

The receiver rotates its surveillance beam periodically among the directions
 $\mathbf{\Phi} = \left\{\phi_1,\phi_2,\cdots,\phi_{\mathrm{Q}}\right\}$, as in Fig. \ref{fig:02_coordinate system}. The duration that the surveillance beam stops at one direction is $\mathrm{T_b}$, and the sweeping period that the surveillance beam traverses all the directions is $\mathrm{T_d} =\mathrm{Q}\mathrm{T_b}$. 

To facilitate the simultaneous localization and tracking, it is assumed that the receiver has collected the receive signals of $J$ ($J\geq 1$) trajectories, respectively. In the remaining of this section, the motion and signal models of the $j$-th trajectory ($j=1,2,\cdots,J$) will be elaborated. 

\subsection{Motion Model}

Since the sweeping period is very short, it can be approximated that the target moves with a constant velocity in a sweeping period. Denote the initial position of the moving target as $\boldsymbol{p}_{j,1} = [x_{j,1}, y_{j,1}]^\mathsf{T}$, and the starting position and velocity of the moving target in the $j$-th trajectory and the $k$-th sweeping period as $\boldsymbol{p}_{j,k} = [x_{j,k}, y_{j,k}]^\mathsf{T}$ and $\boldsymbol{v}_{j,k} = [{v}_{j,k}^x, {v}_{j,k}^y]^\mathsf{T}$, respectively, where $[\cdot ]^{\mathsf{T}}$ denotes the vector (matrix) transpose. The trajectory of the moving target can then be expressed as
\begin{align}
\label{eq:motion_model}
    \boldsymbol{p}_{j,k} & = \boldsymbol{p}_{j,k-1} + \boldsymbol{v}_{j,k-1}\mathrm{T_d}  \\ \nonumber
   & = \boldsymbol{p}_{j,1} + \sum_{n=1}^{k-1}\boldsymbol{v}_{j,n}\mathrm{T_d}, 
   \ k=2,3,\cdots,K_j,
\end{align}
where $K_j$ is the total number of sweeping periods of the $j$-th trajectory. Moreover, the AoA $\phi_{j,k}$ of the echo signal scattered off the target can be written as
\begin{align}
    \phi_{j,k} = \mathrm{atan2}(y_{j,k},x_{j,k}),
    \label{eq: obs_aoa}
\end{align}
where $\mathrm{atan2}(\cdot, \cdot)$ denotes the four-quadrant inverse tangent.

Given the above motion model, the bistatic Doppler frequency of the $m$-th surveillance channel ($m=1,2$) is given by
\begin{align}
    f_{j,k}^{m} &=& - \frac{1}{\lambda_m}
    \left ( 
    \dfrac{\boldsymbol{p}_{j,k}-\boldsymbol{p}_{m}}{\Vert \boldsymbol{p}_{j,k}-\boldsymbol{p}_{m} \Vert} + 
    \dfrac{\boldsymbol{p}_{j,k}-\boldsymbol{p}_{r}}{\Vert \boldsymbol{p}_{j,k}-\boldsymbol{p}_{r} \Vert}
    \right ) ^{\mathsf{T}}
    \boldsymbol{v}_{j,k},
    \label{eq: obs_fd}
\end{align}
where $\lambda_m$ is the carrier wavelength of the signal from the $m$-th transmitter. As a result, given the $j$-th trajectory, denoted as $\mathbf{T}_j=[\boldsymbol{p}_{j,1}, \boldsymbol{v}_{j,1}, \cdots, \boldsymbol{v}_{j,K_j}]$, and the locations of the two transmitters, the ground truth of the AoAs and bistatic Doppler frequencies of all sweeping periods are given by \eqref{eq: obs_aoa} and \eqref{eq: obs_fd}, respectively. Note that the AoAs and bistatic Doppler frequencies are measured via passive sensing, the difference between the ground truth and the measurements can be evaluated. The mmAlert will tune the transmitters' locations and the trajectories, such that the corresponding AoAs and bistatic Doppler frequencies best fit the measurements. 

\subsection{Uplink Transmission Model}

In the $k$-th sweeping period ($\forall{k}$), let $x_{k,q}^{m,j}(t)$, $t\in[0,\mathrm{T_b}]$, be the baseband signal of $m$-th transmitter ($m = 1,2$) when the surveillance beam is at the direction $\phi_q$ ($q = 1,2,\cdots,\mathrm{Q}$), the corresponding receive signal via the reference beam $m$ can be written as
\begin{equation}
    r_{k,q}^{m,j}(t) = h_{k,q}^{m,j}x_{k,q}^{m,j}(t-\tau_{k,q}^{m,j}) + n_{k,q}^{m,j}(t),\ \  0 \leq t \leq \mathrm{T_b},
\end{equation}
where $h_{k,q}^{m,j}$ and $\tau_{k,q}^{m,j}$ denote the baseband channel gain and delay of the LoS path (reference channel) respectively, $n_{k,q}^{m,j}(t)$ denotes the superposition of noise and NLoS echoes. The power of $h_{k,q}^{m,j}x_{k,q}^{m,j}(t-\tau_{k,q}^{m,j})$ is usually much stronger than that of $n_{k,q}^{m,j}(t)$.

Similarly, the receive baseband signal of the surveillance beam from the $m$-th transmitter includes the echo signals from the moving target and static scattering clusters, which can be written as
\begin{align}
	\begin{aligned}
		s_{k,q}^{m,j}(t) &=
		\tilde{h}_{k,q}^{m,j}x_{k,q}^{m,j}\left(t-\tilde{\tau}_{k,q}^{m,j}\right)e^{j2\pi f_{j,k}^{m}t}+\\
		&\sum_{l=1}^{\mathrm{L}_{k,q}^{m,j}}h_{l}x_{k,q}^{m,j}(t-\tau_{l})
		+\tilde{n}_{k,q}^{m,j}(t),\ \  0 \leq t \leq \mathrm{T_b},
		\label{eq: ys}
	\end{aligned}
\end{align} 
where $\tilde{h}_{k,q}^{m,j}$, $\tilde{\tau}_{k,q}^{m,j}$ and $f_{j,k}^{m}$ \vspace{2pt} denote the quasi-static baseband channel gain, delay and Doppler frequency of the surveillance channel, respectively. Moreover, $\mathrm{L}_{k,q}^{m,j}$ is the number of paths scattered from static clusters, $h_{l}$ and $\tau_{l}$ are the complex gain and delay of the $l$-th one, $\tilde{n}_{k,q}^{m,j}(t)$ denotes the noise. As a remark, if the target is not at the direction $\phi_q$ of the receiver, the channel gain $\tilde{h}_{k,q}^{m,j}$ would be of small magnitude. Moreover, the signal from the LoS path may also be received by the surveillance beam, which can be treated as a special static scattering cluster in the second term of the above equation.

Both receive signals of reference and surveillance beams are sampled at the baseband with a period $\mathrm{T_s}$, respectively. The sampled signals can be expressed by
$$
    r_{k,q}^{m,j}[n] = r_{k,q}^{m,j}(n\mathrm{T_s}) \quad \mathrm{and} \quad
    s_{k,q}^{m,j}[n] = s_{k,q}^{m,j}(n\mathrm{T_s}),
$$
where $n = 1,2,\cdots,\mathrm{T_b}/\mathrm{T_s}$. Note that the second term in right hand side of \eqref{eq: ys}, which is of zero Doppler frequency, may interfere the estimation of the target Doppler frequency $f_{j,k}^{m}$. The least-square-based (LS-based) clutter cancellation method elaborated in \cite{tan2005passive} is applied for suppressing the above interference. Denote the surveillance signal after clutter cancellation as $\tilde{s}_{k,q}^{m,j}[n]$ ($\forall k,q,m,j$).

\section{Doppler Frequency \& AoA Estimation} \label{sec:Doppler-AoA}
In this section, the estimation of bistatic Doppler frequency and AoA for the $j$-th trajectory in the $k$-th sweeping period, $\forall j, k,$ is elaborated.
The bistatic Doppler frequency is estimated according to the cross-ambiguity function (CAF) between the reference signal $r_{k,q}^{m,j}[n]$ and surveillance signal $\tilde{s}_{k,q}^{m,j}[n]$. Particularly, the CAF of the receive signals from the $m$-th transmitter in the $k$-th sweeping period at the $q$-th surveillance beam’s direction is defined as
\begin{align}
        R_{j,k}^{m}(q,f)= 
        \max \limits_{\tau}\sum_{n=1}^{N} \left| \tilde{s}_{k,q}^{m,j}[n]
            \left \{r_{k,q}^{m,j}[n-\tau] \right \}^{*}e^{-j2\pi fn\mathrm{T_s}} \right|
	\label{eq: CAF}
\end{align}
where $\{\cdot\}^{*}$ is the complex conjugate, $|\cdot|$\vspace{2pt} is the magnitude, $N =\mathrm{T_b}/\mathrm{T_s}$ denotes the number of samples when the surveillance beam stops at one direction. Since we only focus on the detection of Doppler frequency, the delay $\tau$ is not considered as a parameter of the CAF. In fact, when the signal bandwidth and sampling frequency are low, the delay between $r_{k,q}^{m,j}[n]$ and $\tilde{s}_{k,q}^{m,j}[n]$ is negligible. It can be observed that there would be a peak value of $R_{j,k}^{m}(q,f)$ when $f=f^m_{j,k}$. 

As a remark, note that the AoAs of both surveillance channel 1 and 2 are identical. Denote $\tilde{f}_{j,k}^{m}$ and $\tilde{\phi}_{j,k}$ as the estimated bistatic Doppler frequency and AoA in the $k$-th sweeping period, we have
\begin{align}
    (\tilde{q}_{j,k}^{m}, \tilde{f}_{j,k}^{m})
	= \ \mathop{\arg\max}_{q,f}
	R_{j,k}^{m}(q,f),
	\label{eq:CFAR}
\end{align}
and $\tilde{\phi}_{j, k} = \phi_{\tilde{q}_{j, k}^{m}}$\vspace{2pt}.

As a result, the measurement vector $\boldsymbol{z}_{j,k}$ in the $k$-th sweeping period is defined as
\begin{equation}
    \boldsymbol{z}_{j,k} = [\tilde{\phi}_{j,k},\tilde{f}_{j,k}^{1},\tilde{f}_{j,k}^{2}]^{\mathsf{T}}.
    \label{eq:obs_mea}
\end{equation}
The measurement matrix of the $j$-th trajectory is then given by
\begin{equation}
    \mathbf{Z}_j = [\boldsymbol{z}_{j,1}, \boldsymbol{z}_{j,2}, \cdots, \boldsymbol{z}_{j,K_j}],
    \label{eq:obs_matrix}
\end{equation}
where $K_j$ is the total number of sweeping periods in the $j$-th trajectory. The aggregation of the measurement matrices of all the $J$ trajectories is given by the set
$$
\mathcal{Z} = \{ \mathbf{Z}_1, \mathbf{Z}_2,\cdots, \mathbf{Z}_J \}.$$
The estimated bistatic Doppler frequencies and AoAs of example trajectories according to experiment data will be illustrated in 
Section \ref{sub:estimation}.

As a remark, when a volunteer is considered as the moving target, the movements of different body parts generate distinct Doppler frequency components. In this case, the strongest component can usually represent the torso's motion, due to the maximum radar cross section (RCS) of the torso compared with other body parts. Hence, we can still use the strongest  Doppler frequency component as the detected Doppler frequency. Moreover, the mmAlert proposed in this paper is designed specifically for single-target scenarios. The Doppler frequency and AoA estimation methods employed in this paper are both intended for single-target cases. When multiple targets are present, the AoA-Doppler spectrum in a single sweeping period may exhibit more than one peak. Therefore, to address the multi-target tracking problem, methods such as multi-target tracking (MTT)\cite{MTT} are required to capture the temporal AoA-Doppler characteristics of different targets.
Although the proposed mmWave integrated sensing and communication system has the potential to support multi-target sensing, this falls beyond the scope of the algorithm presented in this paper.

\section{SLAT Problem Formulation} \label{sec:Problem Formulation}
In this section, we shall formulate the simultaneous localization and tracking design as an optimization problem. For the notation convenience, we first define the $j$-th trajectory as matrix 
$$\mathbf{T}_j = [\boldsymbol{p}_{j,1},\boldsymbol{v}_{j,1}, \boldsymbol{v}_{j,2}, \cdots, \boldsymbol{v}_{j,K_j}],$$
and the aggregation set of all the $J$ trajectories is given by
$$\mathcal{T} = \{\mathbf{T}_1, \mathbf{T}_2,\cdots, \mathbf{T}_j\}.$$

Given the two transmitters' locations $\mathcal{M} = \{ \boldsymbol{p}_1, \boldsymbol{p}_2 \}$ and the $j$-th trajectory $\mathbf{T}_j$, the AoAs and the bistatic Doppler frequencies of the two surveillance channels in all the sweeping periods are all determined. Hence, denote the aggregation matrix of actual AoAs and bistatic Doppler frequencies of both surveillance channels given $\mathbf{T}_j$ and $\mathcal{M}$ as
\begin{equation}
\mathbf{H}(\mathbf{T}_j, \mathcal{M}) = 
\left[
\begin{array}{@{}c@{\hspace{1.5em}}c@{\hspace{1.5em}}c@{\hspace{1.5em}}c@{}}
\phi_{j,1} & \phi_{j,2} & \cdots & \phi_{j,K_j} \\
f_{j,1}^1 & f_{j,2}^1 & \cdots & f_{j,K_j}^1 \\
f_{j,1}^2 & f_{j,2}^2 & \cdots & f_{j,K_j}^2
\end{array}
\right],
\label{eq:meaturement_H}
\end{equation}
where $\phi_{j,k}$ and $f_{j,k}^m$ are calculated according to \eqref{eq: obs_aoa} and \eqref{eq: obs_fd}, respectively. 

Note that the AoAs and bistatic Doppler frequencies of all sweeping periods are measured by the receiver. The difference between $\mathbf{H}(\mathbf{T}_j, \mathcal{M})$ and $\mathbf{Z}_j$ ($\forall j$) is the matrix of measurement error of the $j$-th trajectory. Hence, the weighted mean squared error (WMSE) between 
the measurement and the true values given the transmitters' locations $\mathcal{M}$ and the 
the $j$-th trajectory $\mathbf{T}_j$ is defined as
\begin{align}
g_j(\mathbf{T}_j, \mathcal{M}, \mathbf{Z}_j) &\triangleq \frac{1}{K_j} \mathbf{tr} \Biggl\{ \left[ \mathbf{Z}_j - \mathbf{H}(\mathbf{T}_j, \mathcal{M}) \right]^{\mathsf{T}} \mathbf{W} \nonumber \\
&\quad \cdot \left[ \mathbf{Z}_j - \mathbf{H}(\mathbf{T}_j, \mathcal{M}) \right] \Biggr\},
\label{eq:residual_function_full}
\end{align}
where $\mathbf{tr}\{\cdot\}$ denotes the matrix trace, $\mathbf{W} = diag\{w_1,w_2,w_3\}$, $w_1, w_2$ and $w_3$ denote the weights of different features respectively. 

The simultaneous localization and tracking problem can then be conducted by searching the optimal values of $\mathbf{T}_j$ ($\forall j$) and $\mathcal{M}$ such that the summation of WMSE $g_j(\mathbf{T}_j, \mathcal{M}, \mathbf{Z}_j)$ for all trajectories is minimized. Thus,
\begin{align}
\mathcal{P}1: \qquad
\begin{aligned}
         \{\mathcal{T}^{*}, \mathcal{M}^{*}\}= \mathop{\arg \min}_{\{\mathcal{T},\mathcal{M}\}}\sum_{j=1}^{J}g_j(\mathbf{T}_j,\mathcal{M}, \mathbf{Z}_j).
\end{aligned}
\label{eq:cost_func_1}
\end{align}

Since the relations between the AoAs, the bistatic Doppler frequencies and the trajectories, the transmitters' locations are nonconvex, the optimization problem $\mathcal{P}1$ is also nonconvex. Furthermore, when the $J$ trajectories and two transmitters' locations are optimized simultaneously, the number of variables is large, leading to a significant computation complexity. Although the Gaussian Newton method \cite{GN} and the Levenberg-Marquardt (LM) algorithm \cite{LM-1, LM-2} are conventionally employed to solve least squares problems, these methods are characterized by their high computational complexity, when the number of variables is large, and sensitivity to the initial solution. In the following section, a novel low-complexity solution algorithm exploiting the structure of SLAT problem is proposed.

\section{Low-Complexity Solution Algorithm} \label{sec:method}

Note that in problem $\mathcal{P}1$, given the locations of the two transmitters, the search of the $J$ trajectories can be decoupled. This motivates us an alternating optimization framework: given the transmitters' locations, find the optimal $J$ trajectories to minimize the WMSE of the measurements, respectively. Then, given the $J$ trajectories, find the optimal transmitters' locations to minimize the overall WMSE of all trajectories. The above two optimizations are conducted iteratively, which is referred to as the outer iteration. 

Particularly, let $\mathcal{M}^{0}$ be the initial solution of the transmitters' locations, and $\mathcal{M}^{l-1}$ be the solution after the $(l-1)$-th outer iteration, the optimization of the problem $\mathcal{P}1$ in the $l$-th outer iteration, $l=1,2,...$, consists of the following two sub-problems:
\begin{align}
    \mathcal{P}2(l,j): \quad
    \begin{aligned}
         \mathbf{T}_j^l = \mathop{\arg \min}_{\mathbf{T}_j}g_j(\mathbf{T}_j,\mathcal{M}^{l-1},\mathbf{Z}_j), \forall j,
    \end{aligned}
    \label{eq:cossfunction3}
\end{align}
and
\begin{align}
    \mathcal{P}3(l): \quad
    \begin{aligned}
         {\mathcal{M}^l} = \mathop{\arg \min}_{\mathcal{M}}\sum_{j=1}^{J}g_j(\mathbf{T}_j^l,\mathcal{M},\mathbf{Z}_j),
    \end{aligned}
    \label{eq:cossfunction2}
\end{align}
where $\mathbf{T}_j^l$ denotes the $j$-th trajectory after the optimization in the $l$-th outer iteration, respectively. 

According to the geometric relation in \eqref{eqn:tri}, problem $\mathcal{P}_3$ contains only one optimization variable, namely the distance $x_{\mathrm{TX1}}$ between Tx1 and Rx. It can therefore be solved efficiently using a one-dimensional search algorithm. However, the problem $\mathcal{P}2$ is still complicated due to the large number of optimization variables: the initial position of the $j$-th trajectory and the two-dimensional velocities of all sweeping periods should be optimized simultaneously. In fact, the former determines the starting point of the trajectory, and the latter determines its shape. They can be optimized alternately. On one hand, given the trajectory shape, the optimization of starting point consists of only two variables. On the other hand, given the starting point, the existing tracking method, e.g., extended Kalman filter (EKF), can be used to shape the trajectory quickly according to the Doppler and AoA measurements. The above two steps can be conducted iteratively, which is referred to as the inner iteration.

Specifically, denote the operator of shape estimation via the EKF, given the initial position of the trajectory $\boldsymbol{p}_{j,1}$ and measurement matrix $\mathbf{Z}_{j}$, as
\begin{equation}
    \mathbf{\hat{V}}_{j} \triangleq [\boldsymbol{\hat{v}}_{j,1}, \boldsymbol{\hat{v}}_{j,2}, \cdots, \boldsymbol{\hat{v}}_{j,K_j}] = f_{EKF} (\boldsymbol{p}_{j,1}, \mathbf{Z}_j),
    \label{eq: ekf}
\end{equation}
where $\boldsymbol{\hat{v}}_{j,k}$ is the estimated velocity of the $k$-th sweeping period. Let $\boldsymbol{p}_{j,1}^{l,m}$ and $\mathbf{\hat{V}}_{j}^{l,m}$ be the initial position and the shape of the $j$-th trajectory after $m$-th inner iteration in solving problem $\mathcal{P}2(l,j)$. In the $m$-th inner iteration ($m=1, 2, \cdots$), the optimization of problem $\mathcal{P}2(l,j)$ consists of the following two steps:
\begin{equation}
    \mathbf{\hat{V}}_{j}^{l,m} = f_{EKF}(\boldsymbol{p}_{j,1}^{l, m-1}, \mathbf{Z}_j)
    \label{eq: EKF_function}
\end{equation}
and
\begin{align}
    \mathcal{P}4(j,l,m): \\ \nonumber
    \boldsymbol{p}_{j,1}^{l,m} & = \mathop{\arg \min}_{\boldsymbol{p}_{j,1}}g_j([\boldsymbol{p}_{j,1}, \mathbf{\hat{V}}_{j}^{l,m}], \mathcal{M}^{l-1},\mathbf{Z}_j).
    \label{eq: P4}
\end{align}

In the following two parts, the EKF-based trajectory reconstruction in \eqref{eq: ekf} according to the AoA and bistatic Doppler frequency measurements, and the solution algorithm for problem $\mathcal{P}4$ are elaborated, respectively.

\subsection{EKF-based Trajectory Reconstruction}\label{section:EKF}

In this part, the extended Kalman filter for the reconstruction of the $j$-th trajectory according to the initial position $\boldsymbol{p}_{j,1}=[x_{j,1}, y_{j,1}]^\mathsf{T}$ and the measurement matrix $\mathbf{Z}_j = [\boldsymbol{z}_{j,1}, \boldsymbol{z}_{j,2}, \cdots, \boldsymbol{z}_{j,K_j}]$ is elaborated.

First, the state vector of the $j$-th trajectory in the $k$-th sweeping period is defined as
$$\boldsymbol{s}_{j,k} = [x_{j,k}, v_{j,k}^{x}, y_{j,k}, v_{j,k}^{y}]^{\mathsf{T}},$$ 
which consists of the velocity and starting position of this sweeping period.
Hence, the initial state is given by
$$\boldsymbol{s}_{j,1} = [x_{j,1}, v_{j,1}^{x}, y_{j,1}, v_{j,1}^{y}]^{\mathsf{T}},$$ where the initial position $\boldsymbol{p}_{j,1}=[x_{j,1}, y_{j,1}]^\mathsf{T}$ is known, and the initial velocity can be calculated according to $\boldsymbol{p}_{j,1}$ and the Doppler frequency measurement. Particularly,
\begin{align}
        [v_{j,1}^{x}, v_{j,1}^{y}]^{\mathsf{T}} = \mathbf{A}_{f}^{-1} [\tilde{f}^{1}_{j,1},\tilde{f}^{2}_{j,1}]^{\mathsf{T}},
    \end{align}
where 
    \begin{align}
        \mathbf{A}_{f} = 
        \begin{bmatrix}
        -\dfrac{1}{\lambda_1}
        \left ( 
        \dfrac{\boldsymbol{p}_{j,1}-\boldsymbol{p}_{\mathrm{1}}}{\Vert \boldsymbol{p}_{j,1}-\boldsymbol{p}_{\mathrm{1}} \Vert} + 
        \dfrac{\boldsymbol{p}_{j,1}-\boldsymbol{p}_{\mathrm{r}}}{\Vert \boldsymbol{p}_{j,1}-\boldsymbol{p}_{\mathrm{r}} \Vert}
        \right )^{\mathsf{T}}\\[10pt]
        -\dfrac{1}{\lambda_2}
        \left ( 
        \dfrac{\boldsymbol{p}_{j,1}-\boldsymbol{p}_{\mathrm{2}}}{\Vert \boldsymbol{p}_{j,1}-\boldsymbol{p}_{\mathrm{2}} \Vert} + 
        \dfrac{\boldsymbol{p}_{j,1}-\boldsymbol{p}_{\mathrm{r}}}{\Vert \boldsymbol{p}_{j,1}-\boldsymbol{p}_{\mathrm{r}} \Vert}
        \right )^{\mathsf{T}}
        \end{bmatrix}.
        \label{eq:v_calculate}
    \end{align}

Given the above definition of trajectory state, we adopt the following state transition model
\begin{align}
    \boldsymbol{s}_{j,k} = \mathbf{F}\boldsymbol{s}_{j,k-1} + \mathbf{w}_{j,k}, 
\end{align}
where 
\begin{align}
    \mathbf{F} = 
    \begin{bmatrix}
        1 & \mathrm{T_d} & 0 & 0\\
        0 & 1 & 0 & 0\\
        0 & 0 & 1 & \mathrm{T_d}\\
        0 & 0 & 0 & 1
    \end{bmatrix}
\end{align}
is the state transition matrix, and $\mathbf{w}_{j,k} \sim \mathcal{N}(0,\mathbf{Q})$ is the state transition noise as defined in \cite{EKF_noise}, with covariance matrix
\begin{align}
    \mathbf{Q} = 
    \renewcommand{\arraystretch}{1.3} 
    \begin{bmatrix}
        \mathrm{T_d^4}\sigma_{v^x}^2/4 &\mathrm{T_d^3}\sigma_{v^x}^2/2 & 0 & 0\\
        \mathrm{T_d^3}\sigma_{v^x}^2/2 &\mathrm{T_d^2}\sigma_{v^x}^2 & 0 & 0 \\
        0 & 0 & \mathrm{T_d^4}\sigma_{v^y}^2/4 &\mathrm{T_d^3}\sigma_{v^y}^2/2\\
        0 & 0 & \mathrm{T_d^3}\sigma_{v^y}^2/2 &\mathrm{T_d^2}\sigma_{v^y}^2 \\
    \end{bmatrix}.
\end{align}

Moreover, the measurement model, which maps the state to the passive sensing measurements, is given by
\begin{equation}
    \tilde{\boldsymbol{z}}_{j,k} = \boldsymbol{h}(\boldsymbol{s}_{j,k}) + \boldsymbol{u}_{j,k},
    \label{eq:measurement model 1}
\end{equation}
where 
\begin{align}
    \boldsymbol{h}(\boldsymbol{s}_{j,k}) = 
    \begin{bmatrix}
        \mathrm{atan2}(y_{j,k},x_{j,k}) \\
        -\dfrac{1}{\lambda_1}
        \left ( 
        \dfrac{\boldsymbol{p}_{j,k}-\boldsymbol{p}_{1}}{\Vert \boldsymbol{p}_{j,k}-\boldsymbol{p}_{1} \Vert} + 
        \dfrac{\boldsymbol{p}_{j,k}-\boldsymbol{p}_{r}}{\Vert \boldsymbol{p}_{j,k}-\boldsymbol{p}_{r} \Vert}
        \right ) ^{\mathsf{T}}
        \boldsymbol{v}_{j,k} \\
        -\dfrac{1}{\lambda_2}
        \left ( 
        \dfrac{\boldsymbol{p}_{j,k}-\boldsymbol{p}_{2}}{\Vert \boldsymbol{p}_{j,k}-\boldsymbol{p}_{2} \Vert} + 
        \dfrac{\boldsymbol{p}_{j,k}-\boldsymbol{p}_{r}}{\Vert \boldsymbol{p}_{j,k}-\boldsymbol{p}_{r} \Vert}
        \right ) ^{\mathsf{T}}
        \boldsymbol{v}_{j,k}
    \end{bmatrix}
    \label{eq:measurement model 2}
\end{align}
denotes the nonlinear measurement function, $\boldsymbol{u}_{j,k} \sim \mathcal{N}(\mathbf{0}, \mathbf{U})$ is the measurement noise, the covariance matrix $\mathbf{U} = diag(\sigma_{\phi}^{2}, \sigma_{f}^2, \sigma_{f}^2)$.

The trajectory reconstruction via the EKF consists of state prediction and state correction. Let 
$$\boldsymbol{s}_{j,k|k-1}=[x_{j,k|k-1}, v_{j,k|k-1}^{x}, y_{j,k|k-1}, v_{j,k|k-1}^{y}]^{\mathsf{T}}$$
denote the predicted state of the $k$-th sweeping period given the measurement up to the $(k-1)$-th sweeping period, and
$$\boldsymbol{s}_{j,k|k}=[x_{j,k|k}, v_{j,k|k}^{x}, y_{j,k|k}, v_{j,k|k}^{y}]^{\mathsf{T}}$$ 
denote the corrected state of the $k$-th sweeping period given the measurement of the $k$-th sweeping period, $\mathbf{P}_{j,k|k-1}$ and $\mathbf{P}_{j,k|k}$ be the predicted covariance matrix and corrected covariance matrix for $\boldsymbol{s}_{j,k|k-1}$ and $\boldsymbol{s}_{j,k|k}$ respectively. The initial covariance matrix can be given by 
    \begin{align}
        \mathbf{P}_{j,1|1} = 
        \begin{bmatrix}
            \sigma_{x}^{2} & 0 & 0 & 0 \\
            0 & \sigma_{v_x}^{2} & 0 & 0 \\
            0 & 0 & \sigma_{y}^{2} & 0 \\
            0 & 0 & 0 & \sigma_{v_y}^{2} \\
        \end{bmatrix}, 
    \end{align}
where $\sigma_{x}^{2}, \sigma_{v_x}^{2}, \sigma_{y}^{2}$ and $\sigma_{v_y}^{2}$\vspace{2pt} denote the variances of position error and velocity error in the x-coordinate and y-coordinate, respectively.

Note that $\boldsymbol{s}_{j,1|1}=\boldsymbol{s}_{j,1}$, the state prediction for the $k$-th sweeping period can be represented by 
    \begin{align}
        \left \{
        \begin{aligned}
            \boldsymbol{s}_{j,k|k-1} & = \mathbf{F}\boldsymbol{s}_{j,k-1|k-1} \\
            \mathbf{P}_{j,k|k-1} & = \mathbf{F}\mathbf{P}_{j,k-1|k-1}\mathbf{F}^{\mathsf{T}} + \mathbf{Q}
        \end{aligned},
        \quad \forall{k>1}
        \right ..
        \label{eq:EKF_pred}
    \end{align}
Moreover, the state correction can be expressed as 
   \begin{align}
        \begin{cases}
            \boldsymbol{s}_{j,k|k} & = \boldsymbol{s}_{j,k|k-1} + \mathbf{K}_{j,k}(\boldsymbol{z}_{j,k} - h(\boldsymbol{s}_{j,k|k-1}))\\
            \mathbf{P}_{j,k|k} & = (\mathbf{I} - \mathbf{K}_{j,k}\mathbf{C}_{j,k})\mathbf{P}_{k|k-1}
        \end{cases},
        \label{eq:EKF_correction}
    \end{align}
where $\mathbf{I}$ denotes the identity matrix, the near-optimal Kalman gain is written as
\begin{align}
    \mathbf{K}_{j,k} = \mathbf{P}_{j,k|k-1}\mathbf{C}_{j,k}^{\mathsf{T}}(\mathbf{C}_{j,k}\mathbf{P}_{j,k|k-1}\mathbf{C}_{j,k}^{\mathsf{T}} + \mathbf{U})^{-1},
\end{align}
and the Jacobian matrix $\mathbf{C}$ of the measurement function is given by 
\begin{align}
    \mathbf{C}_{j,k} = 
    \begin{bmatrix}
        \dfrac{\partial \phi_{j,k}}{ \partial x_{j,k}} & 0 & \dfrac{\partial \phi_{j,k}}{ \partial y_{j,k}} & 0 \\
        \dfrac{\partial f^{1}_{j,k}}{ \partial x_{j,k}} & \dfrac{\partial f^{1}_{j,k}}{ \partial v^x_{j,k}} & \dfrac{\partial f^{1}_{j,k}}{ \partial y_{j,k}} & \dfrac{\partial f^{1}_{j,k}}{ \partial v^y_{j,k}} \\
        \dfrac{\partial f^{2}_{j,k}}{ \partial x_{j,k}} & \dfrac{\partial f^{2}_{j,k}}{ \partial v^x_{j,k}} & \dfrac{\partial f^{2}_{j,k}}{ \partial y_{j,k}} & \dfrac{\partial f^{2}_{j,k}}{ \partial v^y_{j,k}}
    \end{bmatrix}.
\end{align}
In the above Jacobian matrix, the partial derivatives are given by
\begin{subequations}
    \begin{align}
        \frac{\partial \phi_{j,k}}{ \partial x_{j,k}} & = \frac{-y_{j,k}}{x_{j,k}^2+y_{j,k}^2} 
        \label{eq:par_phi_x}
        \\
        \frac{\partial \phi_{j,k}}{ \partial y_{j,k}} & = \frac{x_{j,k}}{x_{j,k}^2+y_{j,k}^2} 
        \label{eq:par_phi_y}
        \\
        \frac{\partial f^{m}_{j,k}}{ \partial x_{j,k}} & = -\frac{1}{\lambda_m} \left( \frac{(r_{m,j,k}^{y})^2v^x_{j,k} - r_{m,j,k}^{x}r_{m,j,k}^{y}v^y_{j,k}}{(\Vert\boldsymbol{r}_{m,j,k}\Vert)^3} + \right.\\ \nonumber
        & \qquad \qquad \left. \frac{y^2_{j,k}v^x_{j,k}-x_{j,k}y_{j,k}v^y_{j,k}}{(\Vert \boldsymbol{p}_{j,k} \Vert)^3}
        \right)
        \label{eq:par_f_x}
        \\ 
        \frac{\partial f^{m}_{j,k}}{ \partial v^x_{j,k}} & = -\frac{1}{\lambda_m}\left(\frac{r_{m,j,k}^{x}}{\Vert \boldsymbol{r}_{m,j,k} \Vert} + \frac{x_{j,k}}{\Vert \boldsymbol{p}_{j,k} \Vert}
        \right)
        \\
        \frac{\partial f^{m}_{j,k}}{ \partial y_{j,k}} & = -\frac{1}{\lambda_m} \left(\frac{(r_{m,j,k}^{x})^2 v^y_{j,k} - r_{m,j,k}^{x}r_{m,j,k}^{y}v^x_{j,k}}{(\Vert\boldsymbol{r}_{m,j,k}\Vert)^3} + \right.\\ \nonumber
         & \qquad \qquad \left. \frac{x^2_{j,k}v^y_{j,k}-x_{j,k}y_{j,k}v^x_{j,k}}{(\Vert \boldsymbol{p}_{j,k} \Vert)^3}
        \right)
        \label{eq:par_f_y}
        \\
        \frac{\partial f^{m}_{j,k}}{ \partial v^y_{j,k}} & = -\frac{1}{\lambda_m}\left(\frac{r_{m,j,k}^{y}}{\Vert \boldsymbol{r}_{m,j,k} \Vert} + \frac{y_{j,k}}{\Vert \boldsymbol{p}_{j,k} \Vert}
        \right),
    \end{align}
    \label{eq:par_EKF}
\end{subequations}
where 
\begin{align}
    \boldsymbol{r}_{m,j,k} & =[r_{m,j,k}^{x},r_{m,j,k}^{y}]^{\mathsf{T}} =\boldsymbol{p}_{j,k}-\boldsymbol{p}_{\mathrm{TX}m} 
    \nonumber \\ &=[x_{j,k}-x_{\mathrm{TX}m},y_{j,k}-y_{\mathrm{TX}m}]^{\mathsf{T}} \nonumber.
\end{align}
The above state prediction \eqref{eq:EKF_pred} and correction \eqref{eq:EKF_correction} are conducted iteratively, which starts from $k=2$ and update as $k=k+1$ after each iteration. Hence, the overall trajectory can be reconstructed when $k=K_j$. The estimated velocities are expressed as
$$\mathbf{\hat{V}}_{j} = \left[\boldsymbol{\hat{v}}_{j,1},\boldsymbol{\hat{v}}_{j,2},\cdots,\boldsymbol{\hat{v}}_{j,K_j}\right],
$$
where $\boldsymbol{\hat{v}}_{j,k} = \boldsymbol{v}_{j,k|k}, k=1, 2, \cdots, K_j$.

\subsection{Solution Algorithm for Problem $\mathcal{P}4$}
In this part, given the shape of the trajectory $\mathbf{\hat{V}}_{j}$, the transmitters' locations $\mathcal{M}$, and the measurement matrix $\mathbf{Z}_{j}$,
the initial position of the sensing target $\boldsymbol{p}_{j,1}$ is obtained by solving the problem $\mathcal{P}4$, where the Levenberg-Marquard algorithm is adopted. If no confusion arises, the outer and inner iteration indexes $l$ and $m$ in \eqref{eq: P4} are omitted in this part to simplify the notations.

Specifically, let $\mathbf{H}(\boldsymbol{p}_{j,1}) \triangleq \mathbf{H}(\mathbf{T}_j, \mathcal{M})=\mathbf{H}([\boldsymbol{p}_{j,1},\mathbf{\hat{V}}_j], \mathcal{M})$ be the matrix of ground truth in \eqref{eq:meaturement_H} given the trajectory shape $\mathbf{\hat{V}}_j$ and transmitters' locations $\mathcal{M}$, $\mathbf{R}_{j}(\boldsymbol{p}_{j,1}) = \mathbf{Z}_{j}-\mathbf{H}(\boldsymbol{p}_{j,1}) \in \mathbb{R}^{3 \times K_j}$ denote the residual matrix of the $j$-th trajectory, which is a matrix depending on the initial position $\boldsymbol{p}_{j,1}$ in the problem $\mathcal{P}4$, $g_j(\boldsymbol{p}_{j,1})=g_j([\boldsymbol{p}_{j,1}, \mathbf{\hat{V}}_{j}], \mathcal{M},\mathbf{Z}_j) $ be the objective function, the problem $\mathcal{P}4$ is rewritten as 
\begin{align}
      \mathop{\arg \min}_{\boldsymbol{p}_{j,1}} g_{j}(\boldsymbol{p}_{j,1})
      = \mathop{\arg \min}_{\boldsymbol{p}_{j,1}}\frac{1}{K_j} \mathbf{tr}\{\mathbf{R}_{j}^\mathsf{T}(\boldsymbol{p}_{j,1}) \mathbf{W} \mathbf{R}_{j}(\boldsymbol{p}_{j,1})\}.
     \label{eq:cost_func_2}
\end{align}

In order to adopt the LM algorithm, we first introduce the following conclusion on the Jacobian of the residual matrix and the Hessian of the objective function. 

\textit{Lemma 1 : 
Let $vec(\cdot)$ be the column-wise vectorization operator of a matrix, i.e. $vec(\mathbf{R}_{j}) \in \mathbb{R}^{3K_j \times 1}$. The Jacobian $\mathbf{J}_{j}\in \mathbb{R}^{3K_j \times 2}$ of $vec(\mathbf{R}_{j})$ can be written as
\begin{align}
    \mathbf{J}_{j} & = \frac{\partial vec(\mathbf{R}_{j}(\boldsymbol{p}_{j,1}))}{\partial \boldsymbol{p}_{j,1}} = -\dfrac{\partial vec(\mathbf{H}(\boldsymbol{p}_{j,1}))}{\partial \boldsymbol{p}_{j,1}} \\ \nonumber
    & = -\left[\left(\dfrac{\partial \boldsymbol{h}_{j,1}}{\partial \boldsymbol{p}_{j,1}}\right)^{\mathsf{T}}, \left(\dfrac{\partial \boldsymbol{h}_{j,2}}{\partial \boldsymbol{p}_{j,1}}\right)^{\mathsf{T}}, \cdots, \left(\dfrac{\partial \boldsymbol{h}_{j,K_j}}{\partial \boldsymbol{p}_{j,1}}\right)^{\mathsf{T}} \right]^{\mathsf{T}},
    \label{eq:Jacobian}
\end{align}
where $\boldsymbol{h}_{j,k}, k=1,2,\cdots, K_j,$ is the $k$-th column of $\mathbf{H}(\boldsymbol{p}_{j,1})$, and according to \eqref{eq:motion_model},
\begin{align}
\begin{aligned}
   \dfrac{\partial \boldsymbol{h}_{j,k}}{\partial \boldsymbol{p}_{j,1}} =
   \dfrac{\partial \boldsymbol{h}_{j,k}}{\partial \boldsymbol{p}_{j,k}}
   \dfrac{\partial \boldsymbol{p}_{j,k}}{\partial \boldsymbol{p}_{j,1}} =
    \begin{bmatrix}
        \dfrac{\partial \phi_{j,k}}{\partial x_{j,k}}  & \dfrac{\partial \phi_{j,k}}{\partial y_{j,k}} \\
        \dfrac{\partial f^1_{j,k}}{\partial x_{j,k}}  & \dfrac{\partial f^1_{j,k}}{\partial y_{j,k}} \\
        \dfrac{\partial f^2_{j,k}}{\partial x_{j,k}}  & \dfrac{\partial f^2_{j,k}}{\partial y_{j,k}} 
    \end{bmatrix}. 
\end{aligned}
\end{align}
Note that, $\dfrac{\partial \phi_{j,k}}{\partial x_{j,k}}$, $\dfrac{\partial \phi_{j,k}}{\partial y_{j,k}}$, $\dfrac{\partial f^1_{j,k}}{\partial x_{j,k}}$, $\dfrac{\partial f^2_{j,k}}{\partial x_{j,k}}$ , $\dfrac{\partial f^1_{j,k}}{\partial y_{j,k}}$ and $\dfrac{\partial f^2_{j,k}}{\partial y_{j,k}}$\vspace{5pt} are  given by 
\eqref{eq:par_phi_x}, \eqref{eq:par_phi_y},  \eqref{eq:par_f_x} and \eqref{eq:par_f_y}, respectively.}

\textit{
Moreover, the gradient and Hessian matrix of $g_{j}(\boldsymbol{p}_{j,1})$ are given by 
\begin{align}
    \nabla g_j(\boldsymbol{p}_{j,1}) = \dfrac{2}{K_j}\{\mathbf{J}_{j}\}^\mathsf{T}(\mathbf{W} \otimes \mathbf{I}_{K_j}) vec(\mathbf{R}_{j}(\boldsymbol{p}_{j,1})),
    \label{eq:gradient}
\end{align}
and
\begin{align}
    \mathbf{B}_{j} \approx \dfrac{2}{K_j}\mathbf{J}_{j}^{\mathsf{T}}(\mathbf{W} \otimes \mathbf{I}_{K_j})\mathbf{J}_{j},
    \label{eq:Hessian}
\end{align}
where $\otimes$ denotes Kronecker product.}

\textit{Proof:}
The Jacobian in \eqref{eq:Jacobian} can be derived directly from the definition, the gradient \eqref{eq:gradient} and approximate Hessian \eqref{eq:Hessian} follow the conclusion in \cite{LM_Hessian}.

As a result, a suboptimal solution of the problem $\mathcal{P}4$ can be obtained iteratively according to the LM algorithm. Since $g_{j}(\boldsymbol{p}_{j,1})$ is not convex, the iteration may be trapped in not-so-good local optimums. A grid-based sampling of initial candidate solutions is combined with an iterative algorithm to achieve a high-quality suboptimal solution. 

Particularly, let $\mathcal{P}_j \subseteq \mathbb{R}^2$ be the candidate region of the initial solution as  
\begin{align}
\mathcal{P}_j \triangleq \left \{
\boldsymbol{p}_{j,1} \left |
    \begin{aligned}
        | \tilde{\phi}_0 - \phi_{j,1}(\boldsymbol{p}_{j,1}) |   & \le \Delta \phi \\ 
        | \boldsymbol{v}_{j,1} | = \Vert \mathbf{A}_f^{-1}(\boldsymbol{p}_{j,1})\tilde{f}_{j,1} \Vert  & \leq v_{\mathrm{max}}\\
    \end{aligned}
    \right .
    \right \},
\end{align}
where $\Delta \phi$ and $v_{\mathrm{max}}$ denote the width of the narrow beam and the maximum velocity of the target, respectively. The initial solutions, denoted as $\{\boldsymbol{p}_{j,1}^{1,0},\boldsymbol{p}_{j,1}^{2,0}, \cdots, \boldsymbol{p}_{j,1}^{n_,0}\}$, are randomly selected from $\mathcal{P}_j$. 

Thus, the iterative search algorithm for the problem $\mathcal{P}4$ can be written as
\begin{align}
     \boldsymbol{p}_{j,1}^{a,b+1} = \boldsymbol{p}_{j,1}^{a,b} - (\mathbf{B}_{j}(\boldsymbol{p}_{j,1}^{a,b})+\mu\mathbf{I})^{-1} \nabla g_j(\boldsymbol{p}_{j,1}^{a,b}), \\ \nonumber
    \quad a = 1,2,\cdots, n, \quad b = 0,1,2,\cdots,
\end{align}
where $b$ is the iteration index and $\mu$ is the damping factor. 

As a remark, after a number of iterations, some of the intermediate solutions $\{\boldsymbol{p}_{j,1}^{a,b}|\forall a,b\}$ might be trapped in not-so-good local optimums. Discarding these solutions and the following iterations can suppress the computation complexity.

\begin{figure*}[htbp] 
	\centering  
	\includegraphics[width=1\textwidth]{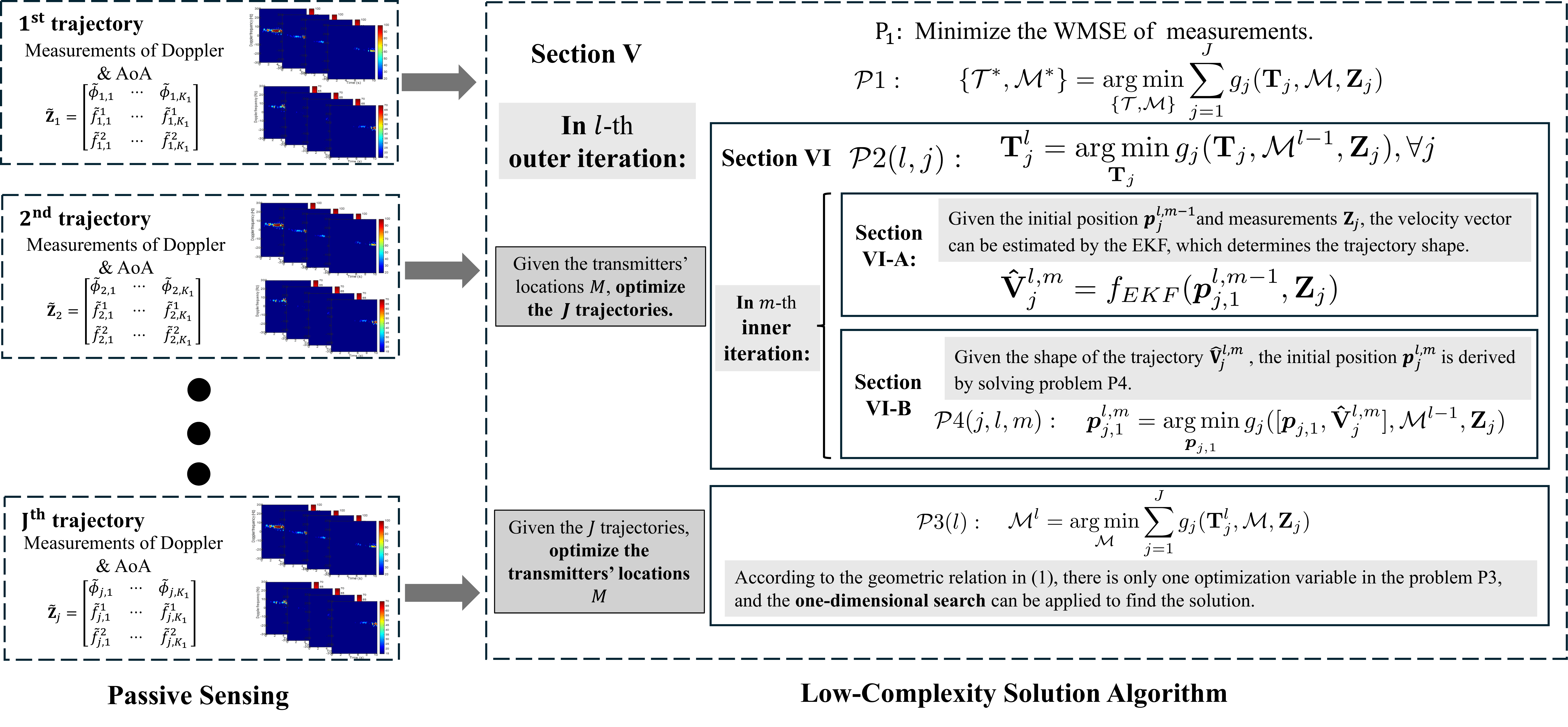}
	\caption{Framework of the proposed low-complexity solution.}
	\label{fig: framework}
\end{figure*}

Finally, let $\boldsymbol{p}_{j,1}^{a,*}$ be the solution after the above iterative algorithm, which is initiated from $\boldsymbol{p}_{j,1}^{a,0}$, and $\tilde{\mathcal{P}}_{j}=\{\boldsymbol{p}_{j,1}^{a,*} | \forall a\}$ be the set of the iteration results, the suboptimal solution of the problem $\mathcal{P}4(j,l,m)$ is given by
\begin{align}
    \hat{\boldsymbol{p}}_{j,1}^{l,m} = \mathop{\arg \min}_{\boldsymbol{p}_{j,1} \in \tilde{\mathcal{P}}_{j}} g_{j}(\boldsymbol{p}_{j,1}).
\end{align}
In summary\vspace{3pt}, the overall low-complexity solution algorithm for the SLAT problem is illustrated in the flowchart of Fig.\ref{fig: framework}.

\section{Experiment Results and Analysis}\label{sec:Experiment}

\begin{figure}[htbp] 
	\centering  
	\includegraphics[width=1\columnwidth]{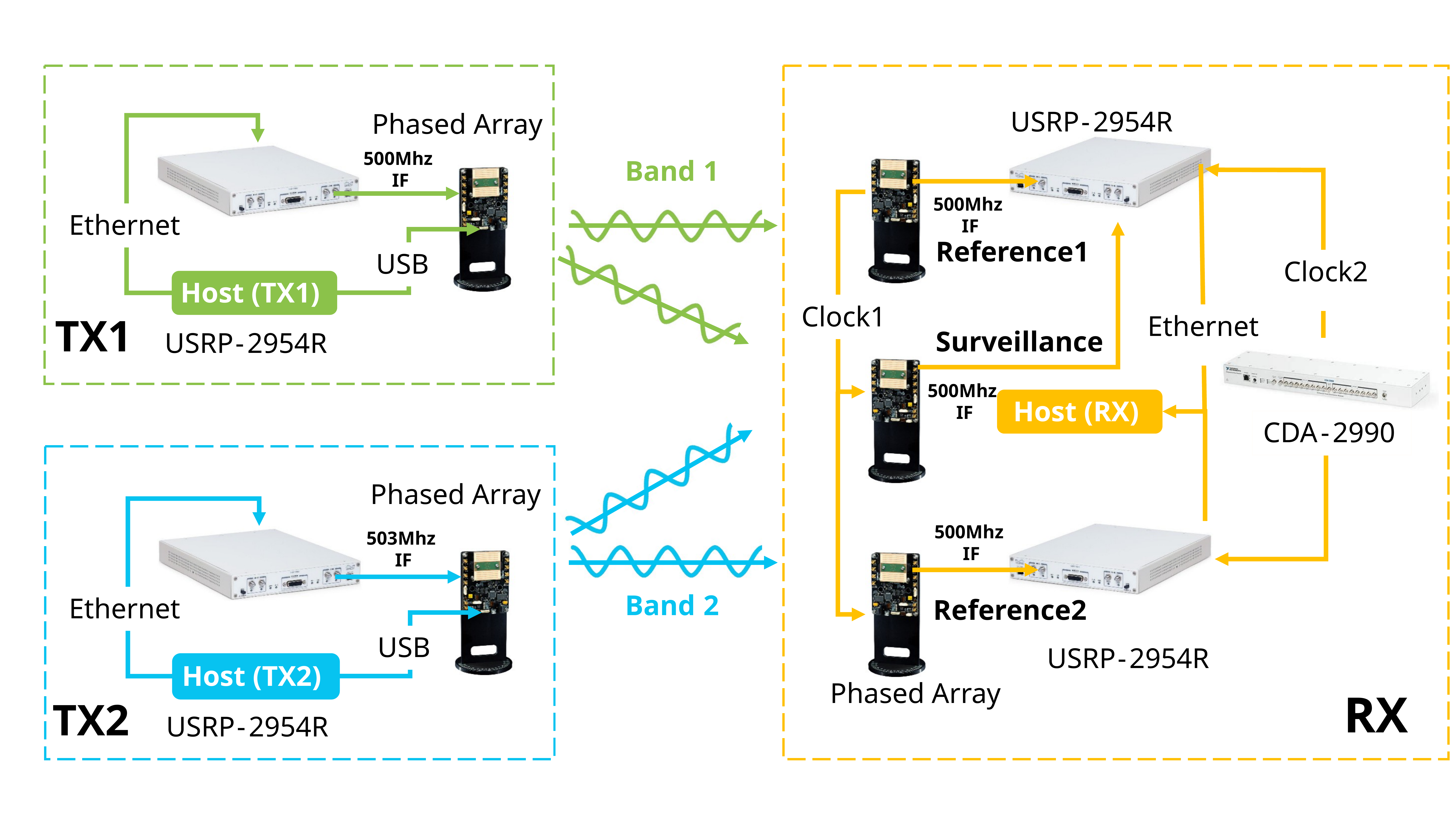}
	\caption{Block diagram of system implementation.}
	\label{fig: system}
\end{figure}

The hardware architecture of the mmAlert system is depicted in Fig. \ref{fig: system}. Each transmitter (transmitter 1 and transmitter 2) consists of an NI USRP-2954R software defined radio (SDR)\cite{NI_USRP_2954} and a Sivers 60GHz phased array \cite{Sivers_EvaluationKits}. The bandwidth of each transmit signal is 1MHz, which consists of a training sequence and an orthogonal frequency division multiplexing (OFDM) modulated data payload. 
The baseband signal of the transmitter 1 is up-converted by the SDR to a central frequency of 500MHz (intermediate frequency, IF). Then, the IF signal is up-converted to 60.98GHz via the Sivers module. Similarly, the central frequency of the IF signal at the transmitter 2 is 503MHz, and the carrier frequency of the RF signal is 60.983GHz. The phased array has 16 transmit antenna elements and 16 receiver antenna elements, respectively, such that both the transmit and receive beams can be adjusted. The transmit beamwidths employed at both transmitters are $90^{\mathrm{o}}$.

The receiver consists of two NI USRP-2954R SDRs and three Sivers 60GHz phased arrays. This is because one SDR can at most connect with two phased arrays. The clocks of the two SDRs are synchronized by CDA-2990 octoclock to ensure consistent carrier frequency offset (CFO). This is important as inconsistent CFOs at different receiver SDRs may introduce significant error in Doppler measurements. Similarly, the three Sivers share the same clock signal in a cascaded manner. At the receiver, two of the phased arrays direct their reference beams to the transmitter 1 and 2, and receive reference channel signals in two distinct frequency bands, respectively. The third phased array provides the surveillance beam, which receives signals of both frequency bands simultaneously. All the receive beamwidths are approximately $10^{\mathrm{o}}$. The surveillance beam switches periodically among $\mathrm{Q}=4$ directions with $\mathbf{\Phi}= \{40.6^{\mathrm{o}}, 28.5^{\mathrm{o}}, 18.5^{\mathrm{o}}, 5.8^{\mathrm{o}}\}$. The surveillance beam stops at one direction with a duration of $\mathrm{T_b}$ = 50 ms, resulting in a complete sweep period of $\mathrm{T_d}$ = $\mathrm{QT_b} = 200$ ms. This enables a velocity detection resolution of 0.025 m/s. The sampling frequency at the receiver is 5MHz, and all the sampled baseband signals are stored in a server for subsequent processing.

The experiment is conducted in an indoor corridor scenario. The moving target can be a Turtlebot robot or a volunteer. The robot's true motion trajectory is recorded by its odometer, while the volunteer's motion trajectory is recorded by a ZED depth camera \cite{zed_camera_webpage}. The timing of both the robot and the ZED camera can be synchronized with the receiver at millisecond level. Since the motion of different parts of the human body are different in walking, we extract the trajectory information of 21 key points of the human body from the depth image data with centimeter-level accuracy, and take the trajectory of shoulder, which is of the same height as the phased array antenna, as the real trajectory of the human body.
In order to analyze the sensitivity of the localization and reconstruction algorithm versus the transmitters' locations, the transmitter 1 is fixed at one position, and the transmitter 2 is placed at three different positions. 50 random trajectories are measured in each scene (each pair of transmitter 1 and 2's positions). Thus, 150 trajectories are measured in total.

\begin{figure*}[htbp] 
  \begin{minipage}{0.49\textwidth}
    \centering
    \includegraphics[width=0.95\linewidth]{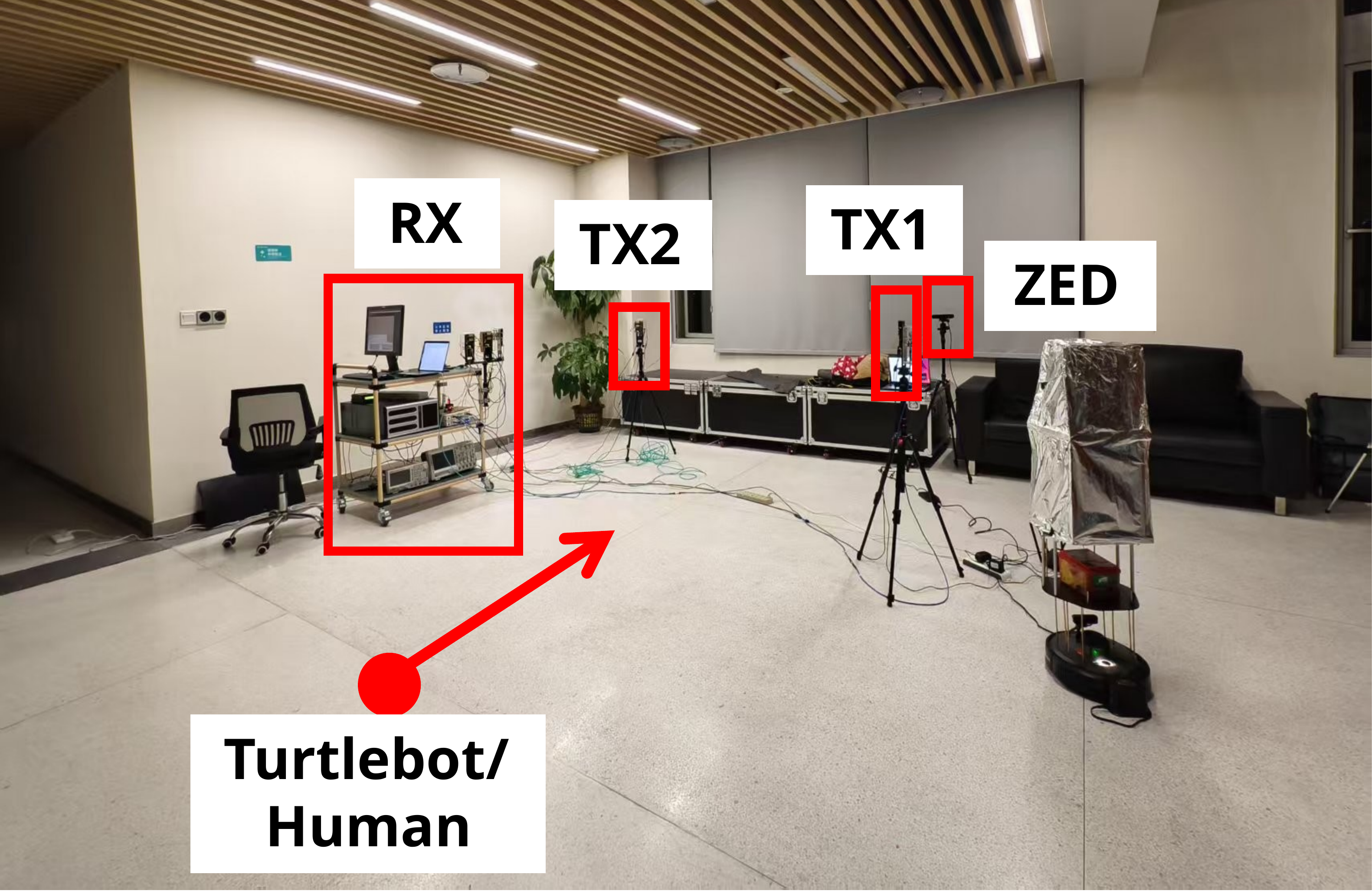}
    \subcaption{}
    \label{fig:exp_a}
  \end{minipage}
  \hfill 
  \begin{minipage}{0.49\textwidth}
    \centering
    \includegraphics[width=0.95\linewidth]{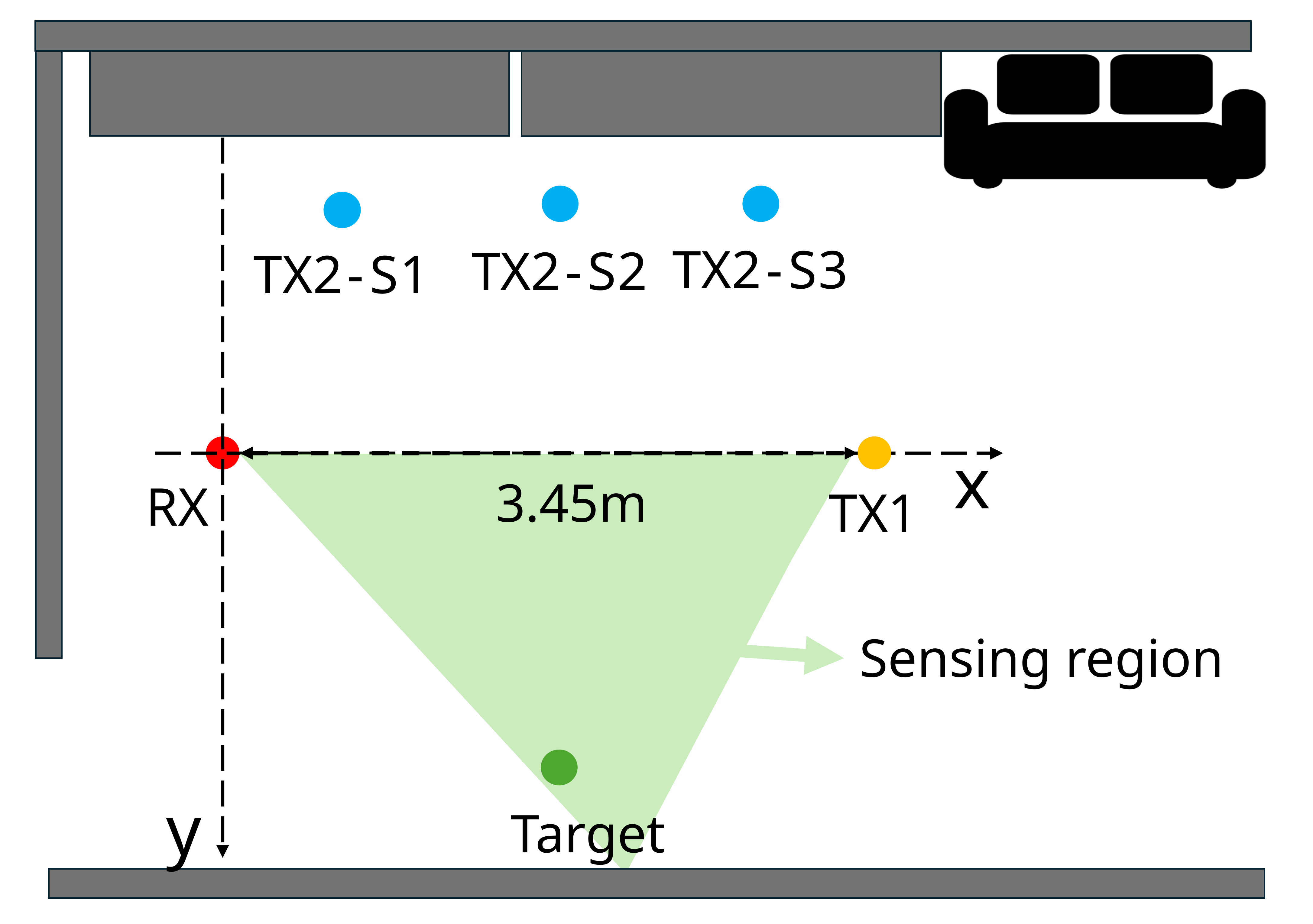}
    \subcaption{}
    \label{fig:exp_b}
  \end{minipage}
    \caption{Illustration of the experiment scenario, where (a) shows a snapshot of the experiment scenario; (b) shows the top view. In (b), the gray area is the obstacle, e.g., the wall, the green area is the sensing region of the system, and the blue circles indicate the transmitter 2's position in the three experiment scenes.}
\end{figure*}

\subsection{Bistatic Doppler Frequency \& AoA Estimation}\label{sub:estimation}

\begin{figure}[htbp]
  \centering
  \begin{subfigure}{\columnwidth}
    \centering
    \includegraphics[width=1\columnwidth]{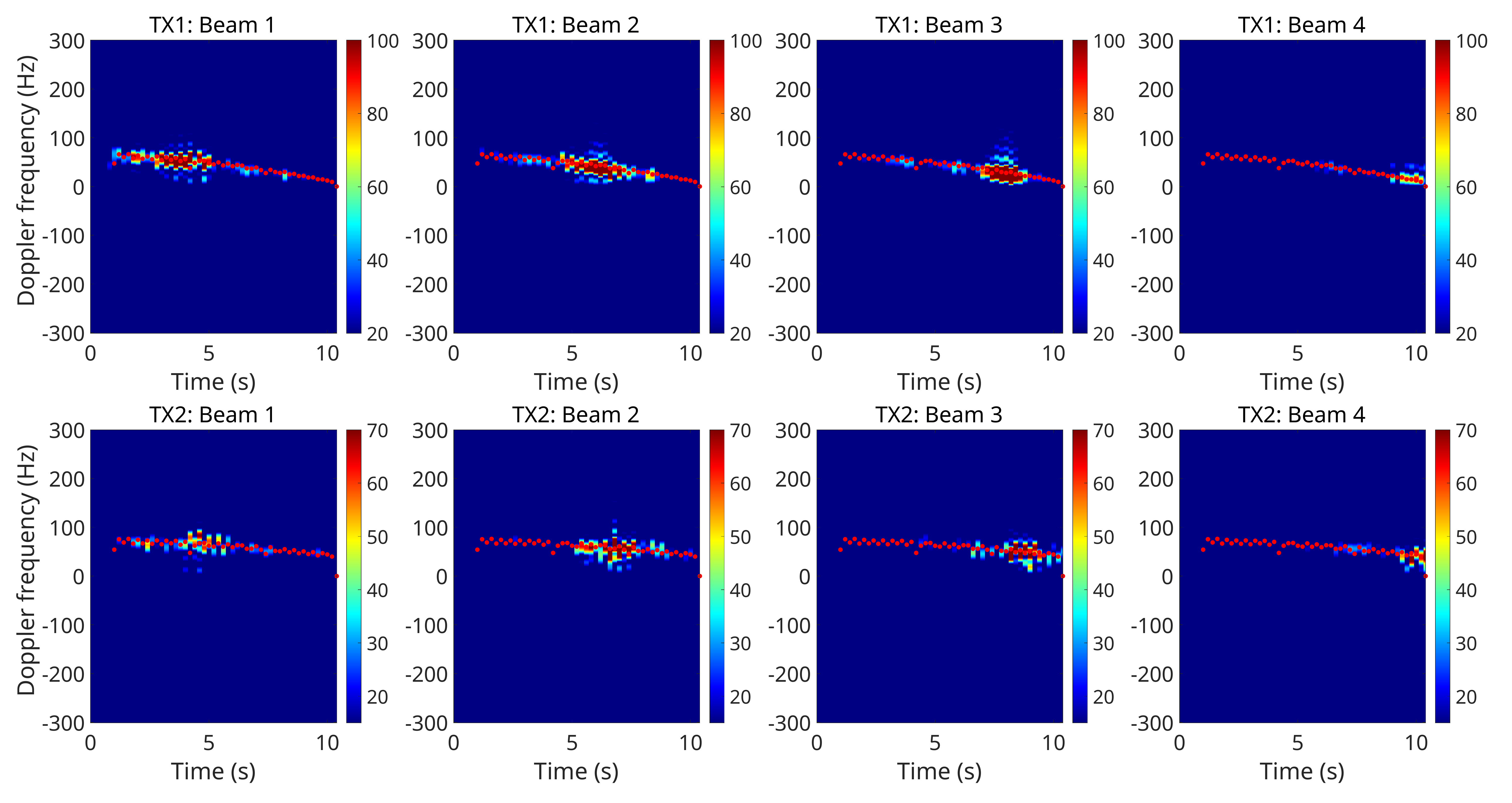}
    \subcaption{Doppler-time spectrum of the surveillance channel 1 and 2 at the four surveillance beam directions.}
    \label{fig:fig6_1a}
  \end{subfigure}
  \begin{subfigure}{0.49\columnwidth}
    \centering
    \includegraphics[width=\linewidth]{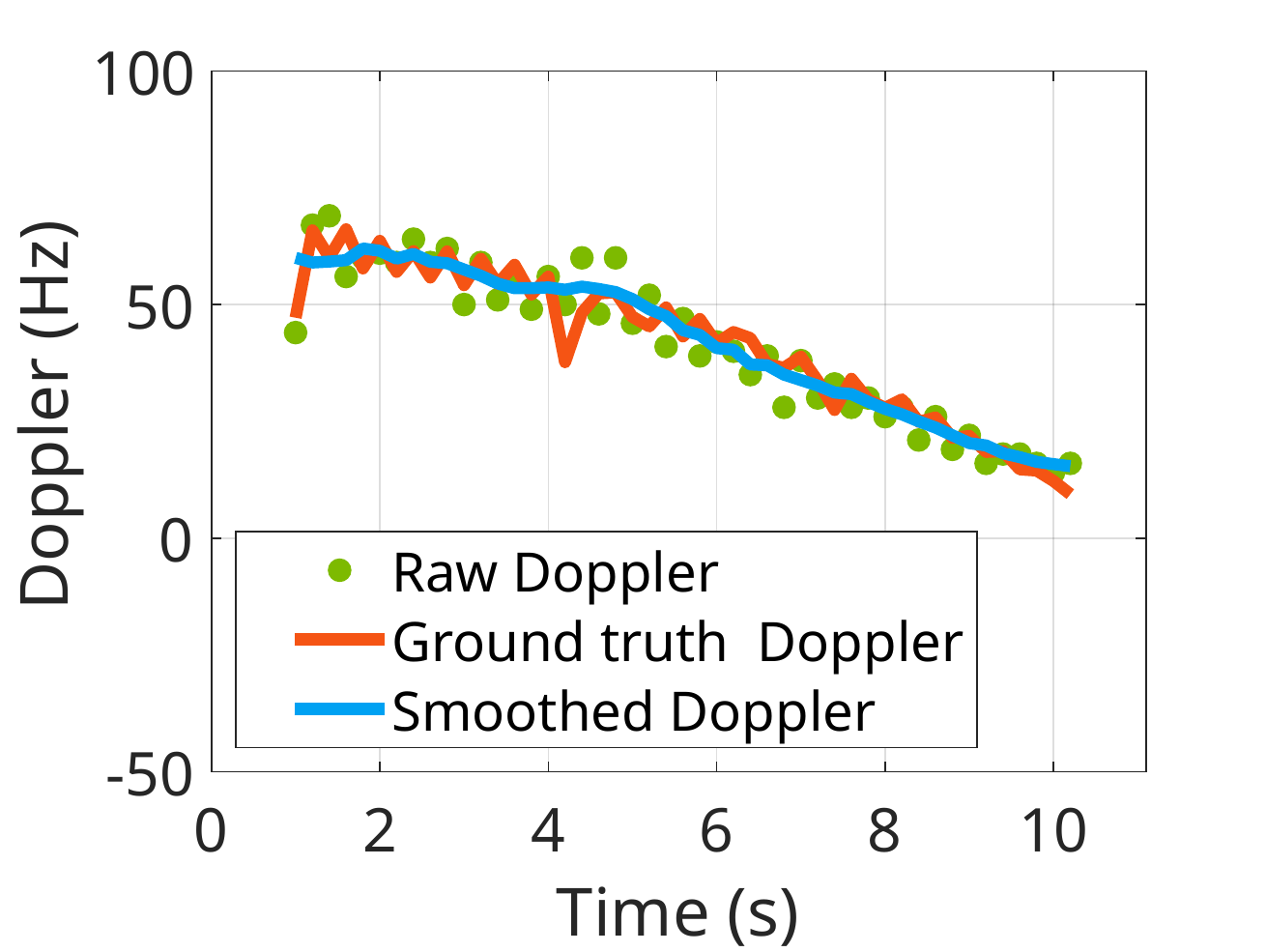}
    \subcaption{Bistatic Doppler frequencies of surveillance channel 1}
    \label{fig:fig6_1b}
  \end{subfigure}
  \hfill
  \begin{subfigure}{0.49\columnwidth}
    \centering
    \includegraphics[width=\linewidth]{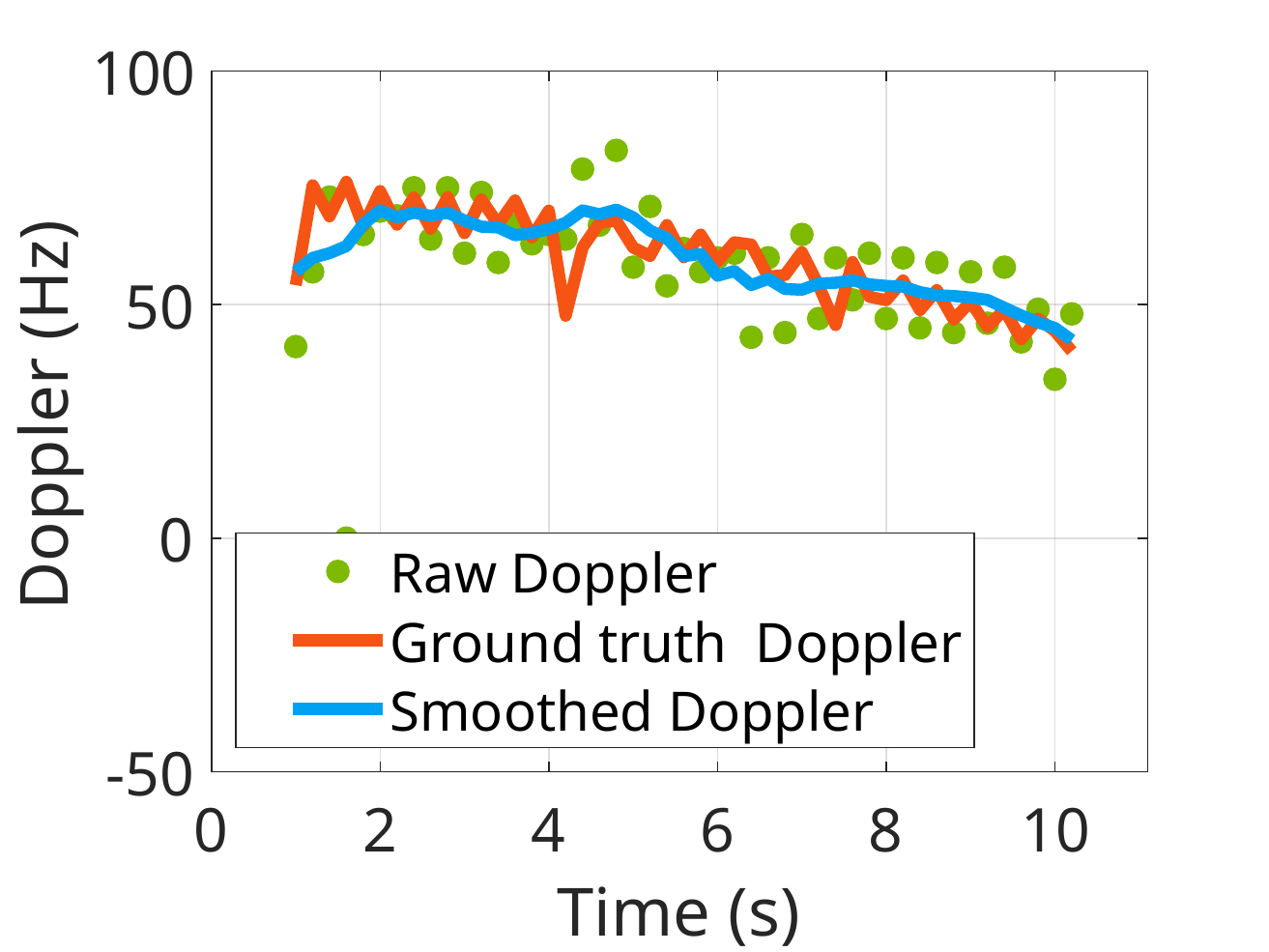}
    \subcaption{Bistatic Doppler frequencies of surveillance channel 2}
    \label{fig:fig6_1c}
  \end{subfigure}
  \hfill
  \begin{subfigure}{0.49\columnwidth}
    \centering
    \includegraphics[width=\linewidth]{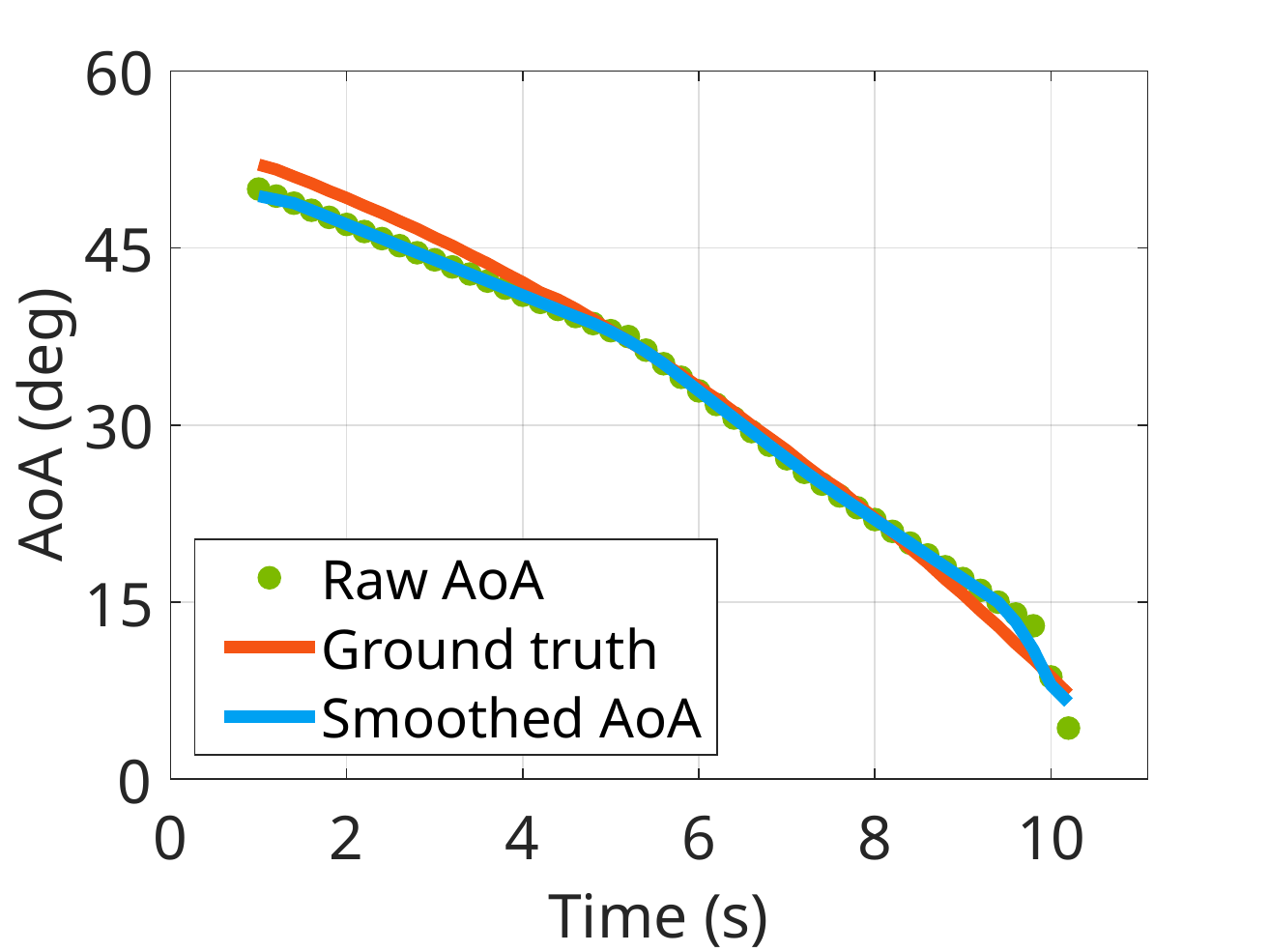}
    \subcaption{AoA}
    \label{fig:fig6_1d}
  \end{subfigure}
\caption{Illustration of Doppler and AoA measurements of one trajectory, where (a) shows the raw Doppler-time plots of surveillance channel 1 and 2; (b) compares the detected bistatic Doppler frequencies before and after smoothing with the ground truth; (c) compares the detected AoAs with the ground truth.}
\label{fig:06_feature}
\end{figure}

In this part, the measurements of bistatic Doppler frequency and AoA are demonstrated.
In Fig. \ref{fig:fig6_1a}, the Doppler-time plots of the two surveillance channels for one example trajectory are demonstrated, respectively, where the transmitter 2 is at TX2-S1 as in Fig. \ref{fig:exp_b}. In this figure, the color intensity represents the significance of the bistatic Doppler frequency component, which is calculated according to the CAF in \eqref{eq: CAF}. Thus, the CAF values of bistatic Doppler frequencies from -300kHz to 300kHz versus time (sweeping period) are illustrated. The first and second rows of the plots correspond to the surveillance channel 1 and 2, respectively. For each surveillance channel, the surveillance beam rotates among 4 different directions, hence, the four plots in each row correspond to the four directions respectively. In each plot, the detected bistatic Doppler frequencies versus time can be observed from the bright regions, and the red line is the ground truth of bistatic Doppler frequency versus time. It can be observed that the moving target is captured by the surveillance beam of different directions in different time periods, which matches the moving trajectory of the target. For example, in this figure, the target is captured by the surveillance beams of the four directions in the time intervals of [1s, 4.8s], [4.8s, 7.4s], [7.4s, 9s] and [9s, 10s], respectively. However, there are still estimation errors in the Doppler measurements.

Therefore, a smoothing algorithm is used to suppress the noise as in Fig. \ref{fig:fig6_1b} and \ref{fig:fig6_1c}, which show the Doppler detection and the smoothed results for the two surveillance channels, respectively. From these two figures, it can be seen that the smoothed bistatic Doppler frequency is closer to the true value.

Due to the $10^{\mathrm{o}}$ beamwidth, the resolution of AoA is worse than that of bistatic Doppler frequency. Consequently, we apply the linear interpolation and smoothing algorithm on the AoA measurements as illustrated in Fig. \ref{fig:fig6_1d}. It is demonstrated that the AoA after linear interpolation and smoothing closely aligns with the ground truth. In the following joint detection, the bistatic Doppler frequencies and AoAs after smoothing are used in the measurement matrices defined in (\ref{eq:obs_matrix}).

\begin{figure}
\centering
  \begin{subfigure}{0.49\columnwidth}
    \centering
    \includegraphics[width=\linewidth]{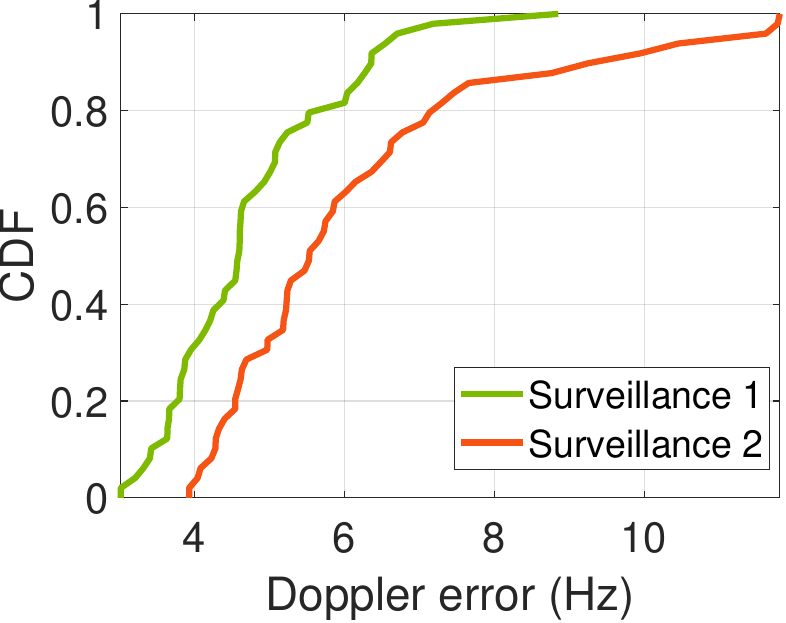}
    \subcaption{Bistatic Doppler frequency}
    \label{fig:fig7_1a}
  \end{subfigure}
  \hfill
  \begin{subfigure}{0.49\columnwidth}
    \centering
    \includegraphics[width=\linewidth]{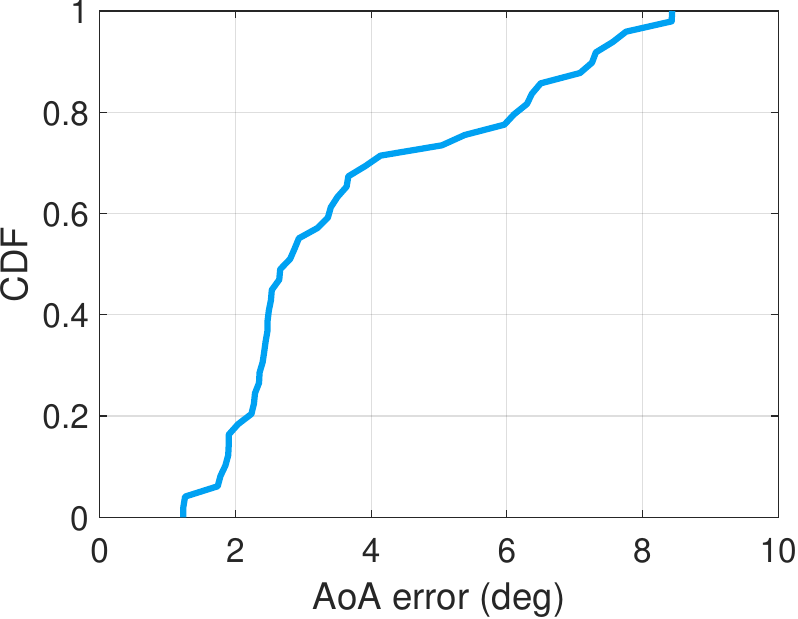}
    \subcaption{AoA}
    \label{fig:fig7_1b}
  \end{subfigure}
\caption{CDFs of measurement errors.}
\label{fig:07_feature_cdf}
\end{figure}

Finally, the cumulative distribution functions (CDFs) of measurement errors for both bistatic Doppler frequency and AoA are shown in Fig. \ref{fig:07_feature_cdf}, where 150 trajectories were measured. The mean errors of bistatic Doppler frequency measurements for the surveillance channel 1 and 2 are 4.7Hz and 6.1Hz, respectively. The mean error of AoA measurements is $3.7^{\mathrm{o}}$. Moreover, bistatic Doppler frequency measurements for the surveillance channel 1 are generally better than that of surveillance channel 2. This is mainly due to their different locations.

\subsection{SLAT with Single trajectory}
\begin{figure}[htbp]
    \centering
    \subfloat[\label{fig:8_1}]{
        \includegraphics[width=0.95\columnwidth]{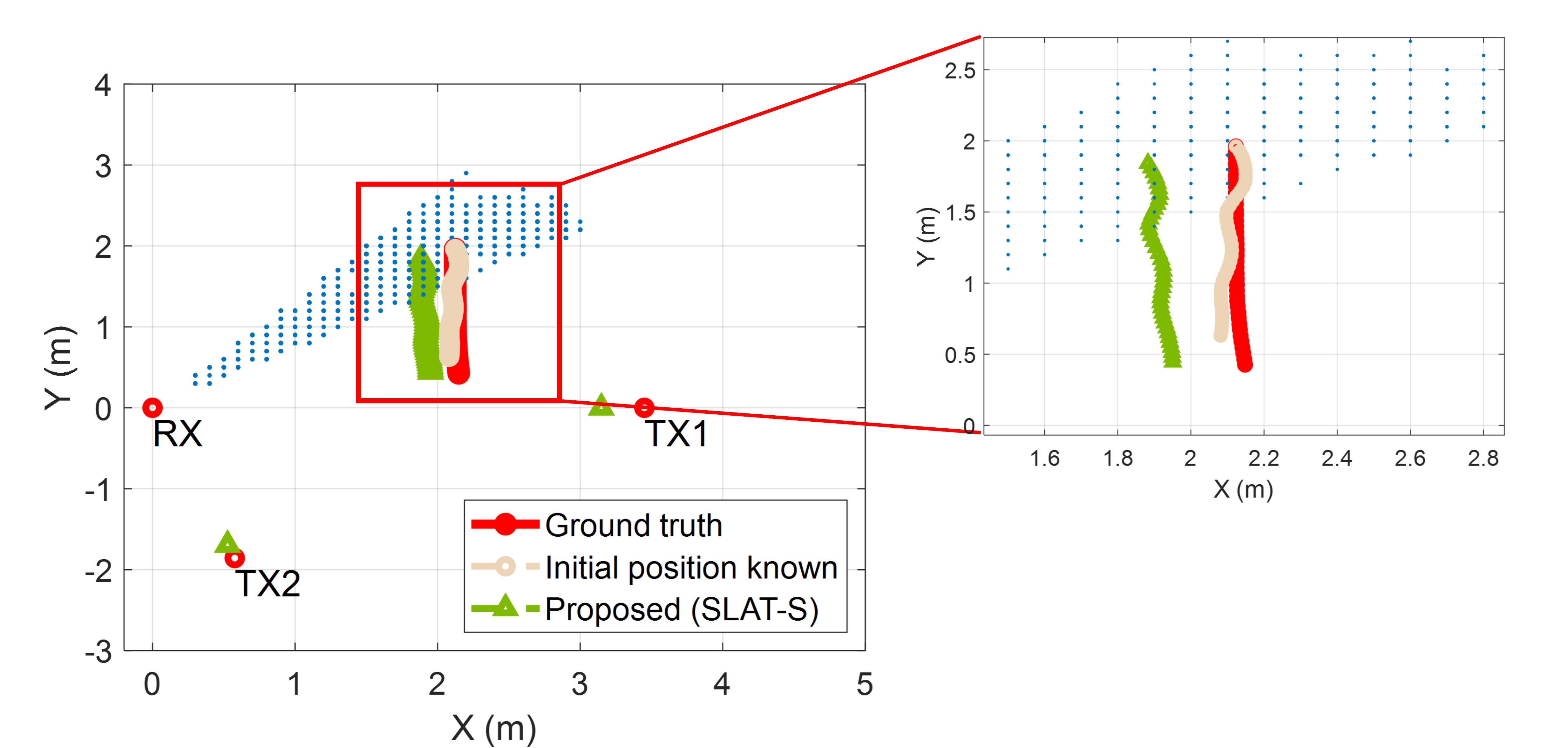}
    }   
    \hfill
    \subfloat[\label{fig:8_2}]{
        \includegraphics[width=0.95\columnwidth]{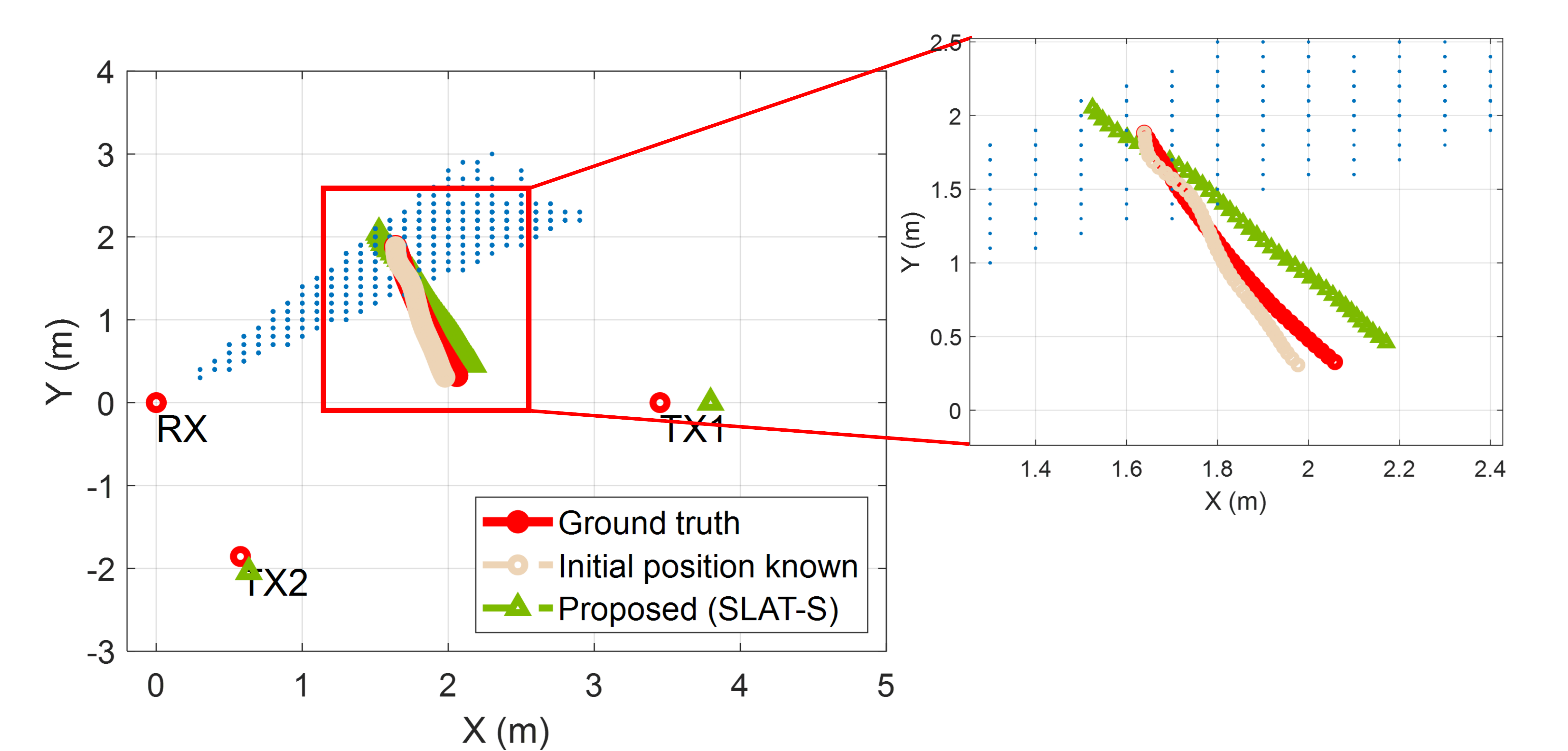} %
    }
    \caption{Two examples of SLAT with single trajectory.}
    \label{fig8:eg_tra_est}
\end{figure}

In this part, a special case, where the mmAlert system works with a single trajectory ($J=1$), is investigated. Particularly, the mmAlert system measures single trajectory of the moving target, then estimates the locations of the two transmitters and the trajectory according to the measurements. The scenario is referred to as SLAT-S for the elaboration convenience. The localization and reconstruction results for two trajectories are illustrated in Fig. \ref{fig:8_1} and \ref{fig:8_2}, where a baseline with perfect knowledge of the two transmitters' locations and the initial position of the trajectory is also plotted for comparison. It can be observed that when the above location knowledge is perfectly known at the receiver, the trajectory reconstruction matches the ground truth very well, where the minor deviations come from the Doppler and AoA measurement errors. On the other hand, without the above knowledge, our proposed SLAT algorithm can still localize the transmitters and reconstruct the trajectory. It can be observed that the localization errors of both transmitters for two examples are 0.31 m, 0.17 m and 0.35 m, 0.2 m, respectively. To evaluate the reconstruction accuracy, the following average Euclidean distance (AED) is adopted as the metric:
\begin{align}
    \mathrm{AED} = \frac{1}{K_j}\sum_{k=1}^{K_j}\Vert\hat{\boldsymbol{p}}_{j,k} - \boldsymbol{p}_{j,k}\Vert,
\end{align}
where $\Vert \cdot \Vert$ denotes the $\mathrm{L}_2$ norm, $\hat{\boldsymbol{p}}_{j,k}$ and $\boldsymbol{p}_{j,k}$ are the reconstructed position and the ground truth at the beginning of the $k$-th sweeping period, respectively. For the 50 trajectories measured when the transmitter 2 is at location TX2-S1, the AED of the two examples are 0.22 m and 0.17 m, respectively.

\subsection{SLAT with Multiple Trajectories}
As discussed in Section \ref{sec:Problem Formulation}, if multiple trajectories are measured by the mmAlert system, the proposed SLAT algorithm is able to suppress the localization errors of both transmitters, and hence, suppress the errors of trajectory reconstruction. It is calculated that when all the 50 trajectories ($J=50$) are jointly used for SLAT (namely SLAT-M), the estimation errors for $x_{\mathrm{TX1}}$, when the transmitter 2 is at TX2-S1, TX2-S2, and TX2-S3, are 0.07 m, 0.16 m and 0.31 m respectively. Note that the locations of the two transmitters can be derived with $x_{\mathrm{TX1}}$. 
Meanwhile, the multi-tone ranging algorithm proposed in \cite{multone} was employed for distance estimation. With a 1~MHz signal bandwidth and a common clock for the transmitter and the receiver (This is hard to implement in real communication systems), this method yields a ranging error of up to 11.6~m, which is substantially larger than the result achieved by mmAlert.

\begin{figure}[htbp]
    \centering
    \subfloat[\label{fig:9_1}]{
        \includegraphics[width=0.48\columnwidth]{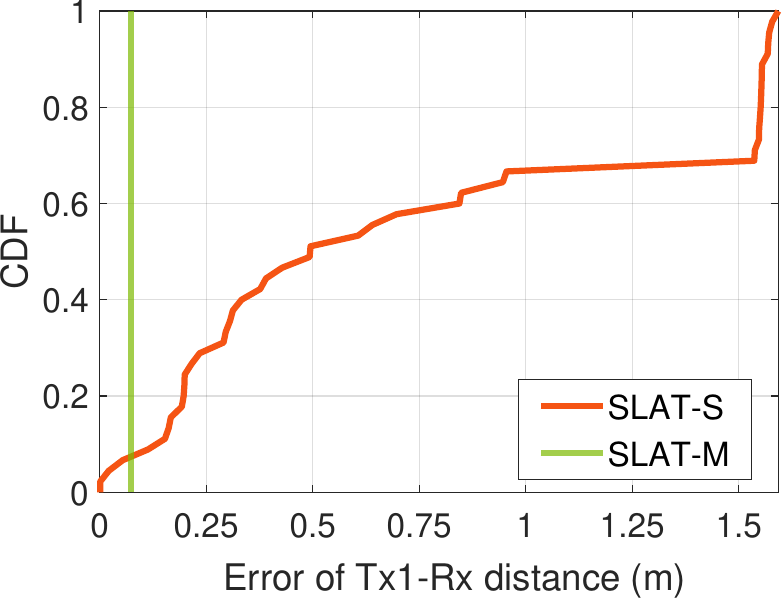} 
    }
    \subfloat[\label{fig:9_2}]{
        \includegraphics[width=0.48\columnwidth]{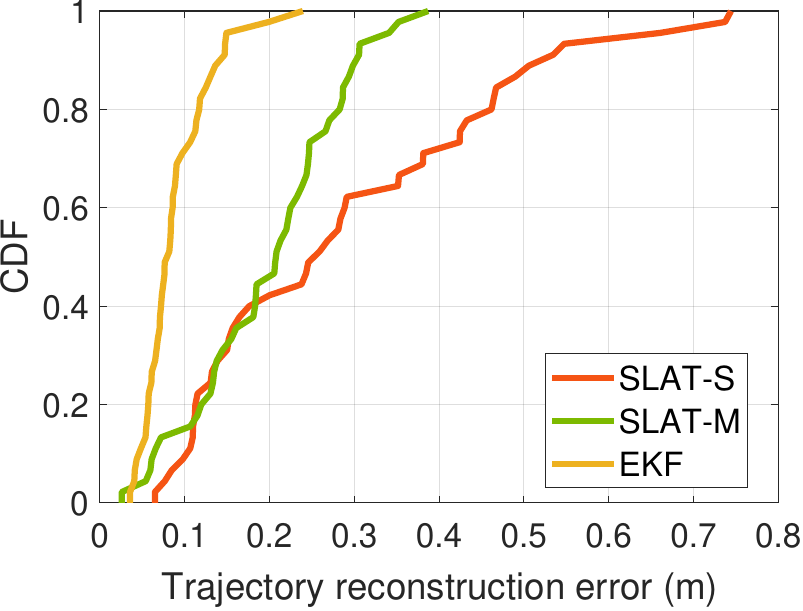} 
    }
    \caption{Comparison of error performance between SLAT-S and SLAT-M.}
    \label{fig9:cdf}
\end{figure}

In Fig. \ref{fig:9_1}, the CDF of detection errors on $x_{\mathrm{TX1}}$, when the transmitter 2 is at location TX2-S1 and SLAT-S is adopted, is plotted. The green line refers to the value of 0.07 m, which is the detection error of SLAT-M. 
Moreover, Fig.~\ref{fig:9_2} presents the CDFs of trajectory reconstruction errors for both SLAT-M and SLAT-S. Meanwhile, the yellow line illustrates the tracking result achieved with the EKF under the condition that the device location and the target's initial position are perfectly known. Hence, the yellow line serves as the performance upper bound. The average trajectory reconstruction error using the SLAT-M algorithm is 0.2~m, meanwhile perfect device location knowledge (yellow line) leads to an error of 0.095 m.
It can be observed that, with SLAT-M, the localization errors of both transmitters are almost negligible. This is significantly better than that of SLAT-S. Moreover, the trajectory diversity is also helpful in suppressing the trajectory reconstruction error. In fact, the transmitters' localization benefits more from multiple trajectories.

\begin{figure}[htbp]
    \centering
    \includegraphics[width=0.7\columnwidth]{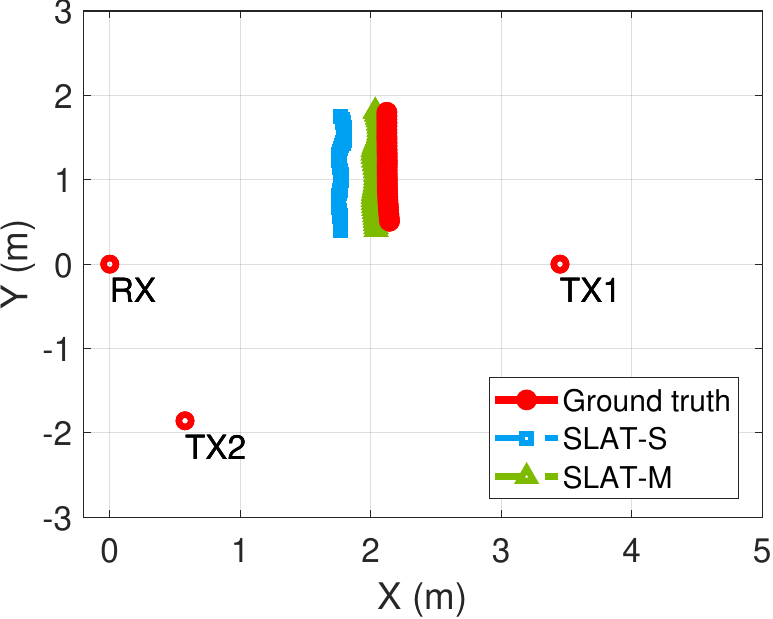}
    \caption{Reconstruction of one trajectory with SLAT-S and SLAT-M.}
    \label{fig:10}
\end{figure}

In Fig. \ref{fig:10}, one reconstructed trajectory is illustrated with the ground truth, where the result of SLAT-S is also plotted for comparison. It can be observed that the reconstructed trajectory via the SLAT-M is much closer to the ground truth than that of SLAT-S.

\subsection{Analysis of Tracking Error}
This part presents a sensitivity analysis of the proposed algorithm through both simulations and experiments. The factors examined include coherent integration time (CIT) $N\mathrm{T_s}$, scanning beam number $\mathrm{Q}$ and beamwidth, transmitter location, and sensing duration $K_j\mathrm{T_d}$. The corresponding results are discussed as follows.

\begin{figure}[htbp]
    \centering
    \subfloat[\label{fig:sim_2}]{
        \includegraphics[width=0.48\columnwidth]{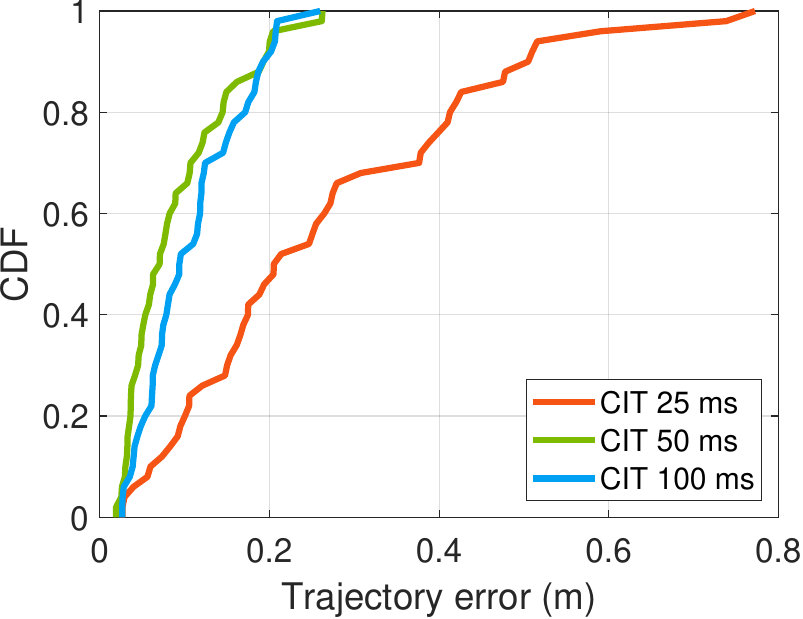}}
    \subfloat[\label{fig:sim_3}]{
        \includegraphics[width=0.48\columnwidth]{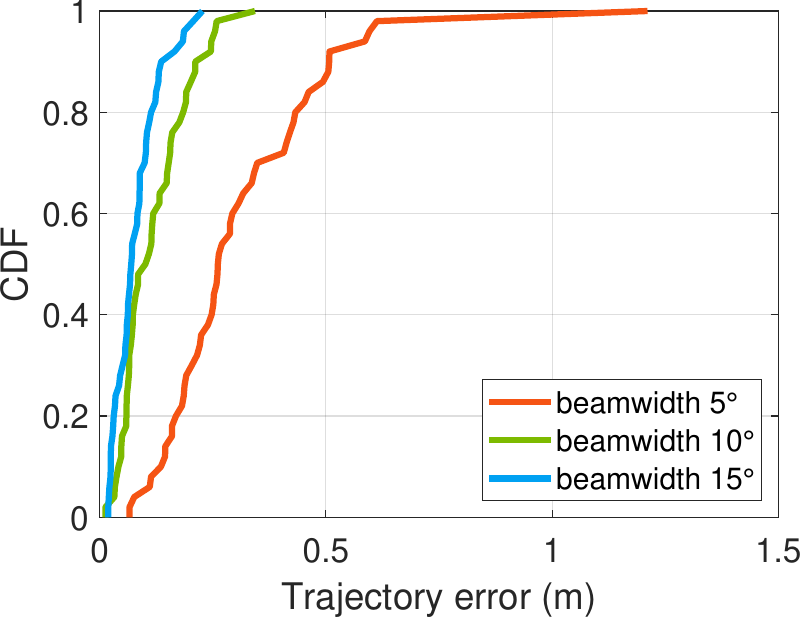}}
    \caption{Simulation comparison of trajectory reconstruction performance with respect to (a) CIT, and (b) beamwidth.}
    \label{fig:sim_analysis}
\end{figure}

\textbf{Impact of CIT.} The Doppler frequency resolution is inversely proportional to the CIT. However, a longer CIT leads to longer sweeping period $\mathrm{T_d}$ and less sweeping periods (smaller $K_j$ for the $j$-th trajectory, $\forall j$). To investigate the impact of CIT on estimation accuracy, we simulated 50 trajectories, where the CIT takes values of 25~ms, 50~ms, and 100~ms, respectively. The results, as shown in the Fig. \ref{fig:sim_2}, indicate that a CIT of 50~ms yields the best performance.

\textbf{Impact of scanning beam number and beamwidth.} We simulated the performance with beamwidths of $5^\circ$, $10^\circ$, and $15^\circ$, respectively. In order to cover the same sensing area, larger beamwidth leads to less sensing direcctions for the surveillance beam (smaller beam number $\mathrm{Q}$) and shorter sweeping period $T_d$. Hence, the corresponding beam numbers are 12, 6, and 4, respectively. The CDFs of trajectory reconstruction errors are shown in the Fig. \ref{fig:sim_3}. Although a $5^\circ$ beamwidth yields a higher AoA estimation accuracy, the update frequency of AoA estimation becomes low, which leads to a significant drop in overall performance. The results for $10^\circ$ and $15^\circ$ beamwidths are comparable, indicating that appropriately widening the beam can improve estimation performance. 

\begin{figure}[htbp]
    \centering
    \subfloat[\label{fig:11_1}]{
    \includegraphics[width=0.48\columnwidth]{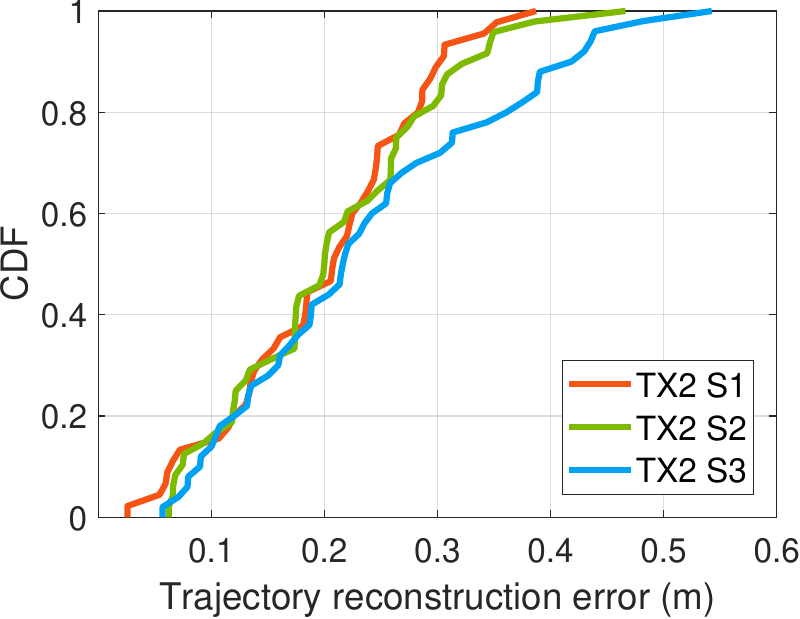}}
    \subfloat[\label{fig:Lentraj}]{
    \includegraphics[width=0.47\columnwidth]{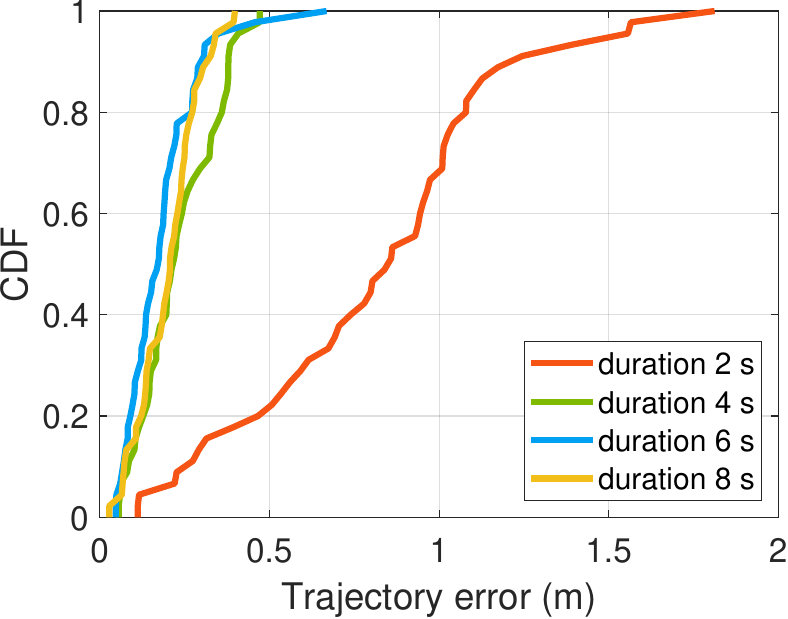}}
    \caption{Experimental comparison of trajectory reconstruction performance with respect to (a) three TX2 positions and (b) sensing duration.}
    \label{fig:11_TX_analysis}
\end{figure}

\textbf{Impact of transmitter location.} In the experiments, the transmitter 2 is placed at the three positions, as shown in Fig. \ref{fig:exp_b}. The CDFs of trajectory reconstruction errors for the three locations with SLAT-M are illustrated in Fig. \ref{fig:11_1}. It can be observed that the worst reconstruction errors for the three scenes are 0.3 m, 0.34 m, and 0.42 m respectively in $90\%$ of the trajectories.

\textbf{Impact of sensing duration.} We obtained observations of trajectories with different lengths by segmenting the original trajectory data from the TX-S1 experimental scenario. The segment durations are 2~s, 4~s, 6~s, and 8~s, respectively. As shown in the Fig. \ref{fig:Lentraj}, when the observation duration is only 2~s, the estimation accuracy degrades significantly due to the limited number of AoA and Doppler observations. However, as the observation duration increases, both localization and trajectory tracking accuracy improve substantially. Moreover, when the observation duration reaches 6~s, the estimation accuracy becomes comparable to that obtained using the full dataset.

\begin{figure}[htbp]
    \centering
    \subfloat[triangular]{\label{fig:triangular}
    \includegraphics[width=0.48\linewidth]{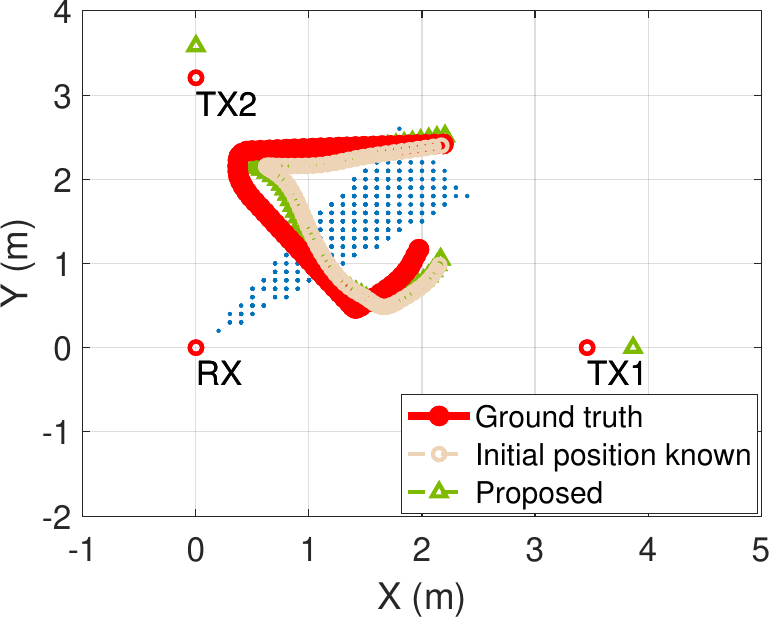}}
    \subfloat[circular]{\label{fig:circular}
    \includegraphics[width=0.48\linewidth]{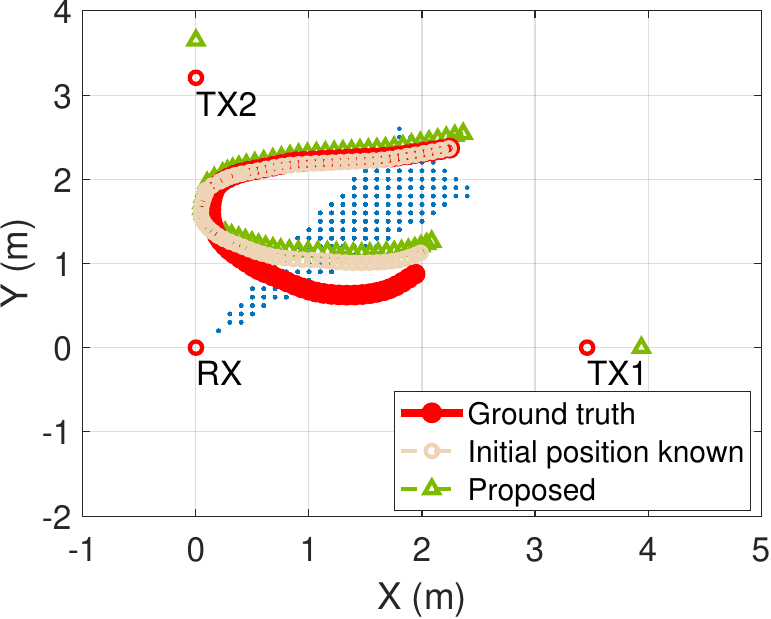}}
    \caption{Comparison of estimation results for two different trajectory types.}
    \label{fig: example for complex trajectory}
\end{figure}

\textbf{Impact of trajectory shape.} We compared the impact of different trajectory shapes on system performance. As indicated by the red circles in Fig. \ref{fig: example for complex trajectory}, the two transmitters Tx1 and Tx2 are deployed along the X-axis and Y-axis, respectively. The estimation results for the triangular trajectory and the circular trajectory are shown by the green lines in Fig. \ref{fig: example for complex trajectory}. Although both trajectories can be successfully reconstructed, the average trajectory estimation errors are 0.15 m and 0.32 m, respectively. This is because the EKF state transition model assumes uniform linear motion of the target between two adjacent sweeping periods. In other words, the triangular trajectory fits the state transition model better than the circular trajectory.

Moreover, as demonstrated in Table \ref{tab:table1}, the average trajectory reconstruction error and average transmitter localization error are presented for the three scenes. The SLAT-M keeps the average trajectory reconstruction error below 0.25 m in all three scenes. The transmitter localization errors with SLAT-M are 0.07 m, 0.16 m, and 0.31 m in the three scenes, respectively. Thus, the accuracy of transmitters' localization depends on the angle between the two transmitters and the target. For example, the position of TX2-S3 is the closest to transmitter 1, leading to the least diversity in the measurement angle. Hence, it has the worst localization and reconstruction errors.

\renewcommand{\arraystretch}{1.5}
\begin{table*}[htbp]
    \centering
    \caption{Average error of trajectory reconstruction and TXs' localization.}
    \label{tab:table1}
    \begin{tabular}{|l|*{6}{c|}}
        \hline
        \multirow{2}{*}{\diagbox[width=10em]{Method}{Scenario}} & \multicolumn{2}{c|}{TX2-S1} & \multicolumn{2}{c|}{TX2-S2} & \multicolumn{2}{c|}{TX2-S3} \\
        \cline{2-7}
        & Trajectories (m) & Transmitters (m) & Trajectories (m) & Transmitters (m) & Trajectories (m) & Transmitters (m) \\
        \hline
        SLAT-S & 0.29 & 0.76 & 0.34 & 0.87 & 0.37 & 0.95 \\
        \hline
        SLAT-M & 0.2 & 0.07 & 0.22 & 0.16 & 0.24 & 0.31 \\
        \hline
    \end{tabular}
\end{table*}

\section{Conclusions}\label{sec:conclusions}
In this paper, we propose mmAlert, to address the simultaneous localization and tracking problem in mmWave communication systems. The mmAlert is implemented in an uplink system with at least two transmitters and one receiver. Without prior knowledge on the transmitters' locations, the proposed method could reconstruct multiple trajectories of a moving target via the passive sensing of the AoAs and bistatic Doppler frequencies, and meanwhile, estimate the locations of the two transmitters. This is because the AoAs and bistatic Doppler frequencies of the moving target depend strongly on the locations of the transmitters. In mmAlert, a low-complexity algorithm is proposed to handle the above joint estimation, which integrates the techniques of alternating optimization and extended Kalman filter. The experimental results demonstrate that mmAlert can achieve a transmitter positioning accuracy of 0.07 m with an average trajectory reconstruction error of 0.2 m when 50 trajectories are utilized. Moreover, the accuracy will decay if only one trajectory is considered in the above joint estimation. This justifies the performance gain due to trajectory diversity. Finally, the proposed method is not limited in the mmWave band, its extension to the frequency bands below 7GHz (sub-7GHz) would be investigated in the future. Moreover, mmAlert can distinguish multiple moving targets by exploiting differences in the Doppler-AoA plane. Multi-target tracking therefore constitutes another promising direction for future research.

\bibliographystyle{IEEEtran}
\bibliography{mmAlert_2}

@article{mmwave,
  title={A survey of millimeter wave communications (mmWave) for 5G: opportunities and challenges},
  author={Niu, Yong and Li, Yong and Jin, Depeng and Su, Li and Vasilakos, Athanasios V},
  journal={Wireless networks},
  volume={21},
  pages={2657--2676},
  year={2015},
  publisher={Springer}
}

@article{mmWave_2,
  author={Li, Jing and Niu, Yong and Wu, Hao and Ai, Bo and Chen, Sheng and Feng, Zhiyong and Zhong, Zhangdui and Wang, Ning},
  journal={IEEE Communications Surveys \& Tutorials}, 
  title={Mobility Support for Millimeter Wave Communications: {Opportunities and Challenges}}, 
  year={2022},
  volume={24},
  number={3},
  pages={1816-1842},
  doi={10.1109/COMST.2022.3176802}}

@article{mmWave_beam,
  title={Beam alignment in mmWave V2X communications: A survey},
  author={Tan, Jingru and Luan, Tom H and Guan, Wenbo and Wang, Yuntao and Peng, Haixia and Zhang, Yao and Zhao, Dongmei and Lu, Ning},
  journal={IEEE Communications Surveys \& Tutorials},
  volume={26},
  number={3},
  pages={1676--1709},
  year={2024},
  publisher={IEEE}
}

@article{mmwave_ISAC,
  title={Active sensing for communications by learning},
  author={Sohrabi, Foad and Jiang, Tao and Cui, Wei and Yu, Wei},
  journal={IEEE Journal on Selected Areas in Communications},
  volume={40},
  number={6},
  pages={1780--1794},
  year={2022},
  publisher={IEEE}
}

@ARTICLE{wifi_tracking1,
  author={Wang, Zhongqin and Zhang, J. Andrew and Xu, Min and Guo, Y. Jay},
  journal={IEEE Transactions on Mobile Computing}, 
  title={Single-Target Real-Time Passive WiFi Tracking}, 
  year={2023},
  volume={22},
  number={6},
  pages={3724-3742},
  doi={10.1109/TMC.2022.3141115}}

@ARTICLE{wifi_tracking2,
  author={Niu, Kai and Wang, Xuanzhi and Zhang, Fusang and Zheng, Rong and Yao, Zhiyun and Zhang, Daqing},
  journal={IEEE Journal on Selected Areas in Communications}, 
  title={Rethinking Doppler Effect for Accurate Velocity Estimation With Commodity WiFi Devices}, 
  year={2022},
  volume={40},
  number={7},
  pages={2164-2178},
  doi={10.1109/JSAC.2022.3155523}}

@article{device_loc1,
  title={DAFI: WiFi-based device-free indoor localization via domain adaptation},
  author={Li, Hang and Chen, Xi and Wang, Ju and Wu, Di and Liu, Xue},
  journal={Proceedings of the ACM on Interactive, Mobile, Wearable and Ubiquitous Technologies},
  volume={5},
  number={4},
  pages={1--21},
  year={2021},
  publisher={ACM New York, NY, USA}
}

@ARTICLE{device_loc2,
  author={Zhang, Xianan and Zhang, Yu and Liu, Guanghua and Jiang, Tao},
  journal={IEEE Transactions on Vehicular Technology}, 
  title={AutoLoc: Toward Ubiquitous AoA-Based Indoor Localization Using Commodity WiFi}, 
  year={2023},
  volume={72},
  number={6},
  pages={8049-8060},
  doi={10.1109/TVT.2023.3243912}}

@ARTICLE{sec1_1,
  author={Wang, Xiong and Kong, Linghe and Kong, Fanxin and Qiu, Fudong and Xia, Mingyu and Arnon, Shlomi and Chen, Guihai},
  journal={IEEE Communications Surveys \& Tutorials}, 
  title={Millimeter Wave Communication: {A Comprehensive Survey}}, 
  year={2018},
  volume={20},
  number={3},
  pages={1616-1653},
  doi={10.1109/COMST.2018.2844322}}

@ARTICLE{sec1_2,
  author={Kumar, Dileep and Kaleva, Jarkko and Tölli, Antti},
  journal={IEEE Transactions on Wireless Communications}, 
  title={Blockage-Aware Reliable mmWave Access via Coordinated Multi-Point Connectivity}, 
  year={2021},
  volume={20},
  number={7},
  pages={4238-4252},
  doi={10.1109/TWC.2021.3057227}}

@ARTICLE{sec1_4,
  author={Gao, Zhen and Wan, Ziwei and Zheng, Dezhi and Tan, Shufeng and Masouros, Christos and Ng, Derrick Wing Kwan and Chen, Sheng},
  journal={IEEE Transactions on Wireless Communications}, 
  title={Integrated Sensing and Communication With mmWave Massive MIMO: A Compressed Sampling Perspective}, 
  year={2023},
  volume={22},
  number={3},
  pages={1745-1762},
  doi={10.1109/TWC.2022.3206614}}

@ARTICLE{sec1_5,
  author={Xue, Qing and Ji, Chengwang and Ma, Shaodan and Guo, Jiajia and Xu, Yongjun and Chen, Qianbin and Zhang, Wei},
  journal={IEEE Communications Surveys \& Tutorials}, 
  title={A Survey of Beam Management for mmWave and THz Communications Towards 6G}, 
  year={2024},
  volume={26},
  number={3},
  pages={1520-1559},
  doi={10.1109/COMST.2024.3361991}}

@ARTICLE{RSSI,
  author={Hoang, Minh Tu and Yuen, Brosnan and Dong, Xiaodai and Lu, Tao and Westendorp, Robert and Reddy, Kishore},
  journal={IEEE Internet of Things Journal}, 
  title={Recurrent Neural Networks for Accurate RSSI Indoor Localization}, 
  year={2019},
  volume={6},
  number={6},
  pages={10639-10651},
  doi={10.1109/JIOT.2019.2940368}}

@INPROCEEDINGS{CSI,
  author={Wang, Xuyu and Gao, Lingjun and Mao, Shiwen and Pandey, Santosh},
  booktitle={2015 IEEE Wireless Communications and Networking Conference (WCNC)}, 
  title={DeepFi: Deep learning for indoor fingerprinting using channel state information}, 
  year={2015},
  volume={},
  number={},
  pages={1666-1671},
  doi={10.1109/WCNC.2015.7127718}}

@inproceedings{AoA,
  title={{ArrayTrack}: A {Fine-Grained} indoor location system},
  author={Xiong, Jie and Jamieson, Kyle},
  booktitle={10th USENIX Symposium on Networked Systems Design and Implementation (NSDI 13)},
  pages={71--84},
  year={2013}
}

@ARTICLE{AoA_2,
  author={Yang, Shuai and Zhang, Dongheng and Song, Ruiyuan and Yin, Pengfei and Chen, Yan},
  journal={IEEE Transactions on Mobile Computing}, 
  title={Multiple {WiFi} Access Points Co-Localization Through Joint {AoA} Estimation}, 
  year={2024},
  volume={23},
  number={2},
  pages={1488-1502},
  doi={10.1109/TMC.2023.3239377}}

@inproceedings{ToF_2,
author = {Kotaru, Manikanta and Joshi, Kiran and Bharadia, Dinesh and Katti, Sachin},
title = {SpotFi: Decimeter Level Localization Using {WiFi}},
year = {2015},
isbn = {9781450335423},
publisher = {Association for Computing Machinery},
address = {New York, NY, USA},
doi = {10.1145/2785956.2787487},
pages = {269–282},
numpages = {14},
location = {London, United Kingdom},
series = {SIGCOMM '15}
}

@ARTICLE{RTT,
  author={Ma, Chengqi and Wu, Bang and Poslad, Stefan and Selviah, David R.},
  journal={IEEE Transactions on Mobile Computing}, 
  title={{Wi-Fi} {RTT} Ranging Performance Characterization and Positioning System Design}, 
  year={2022},
  volume={21},
  number={2},
  pages={740-756},
  doi={10.1109/TMC.2020.3012563}
 }

@inproceedings{MonoLoco,
author = {Soltanaghaei, Elahe and Kalyanaraman, Avinash and Whitehouse, Kamin},
title = {{Multipath Triangulation: Decimeter-level {WiFi} Localization and Orientation with a Single Unaided Receiver}},
year = {2018},
isbn = {9781450357203},
publisher = {Association for Computing Machinery},
address = {New York, NY, USA},
doi = {10.1145/3210240.3210347},
pages = {376–388},
numpages = {13},
location = {Munich, Germany},
series = {MobiSys '18}
}

@ARTICLE{wifi_fingerprint,
  author={Zhu, Xiaoqiang and Qiu, Tie and Qu, Wenyu and Zhou, Xiaobo and Atiquzzaman, Mohammed and Wu, Dapeng Oliver},
  journal={IEEE Transactions on Mobile Computing}, 
  title={{BLS-Location}: A Wireless Fingerprint Localization Algorithm Based on Broad Learning}, 
  year={2023},
  volume={22},
  number={1},
  pages={115-128},
  doi={10.1109/TMC.2021.3073005}}

@inproceedings{Multi_carrier,
  title={{Decimeter-Level} localization with a single {WiFi} access point},
  author={Vasisht, Deepak and Kumar, Swarun and Katabi, Dina},
  booktitle={13th USENIX symposium on networked systems design and implementation (NSDI 16)},
  pages={165--178},
  year={2016}
}

@inproceedings{widar,
  title={Widar: {Decimeter-level} passive tracking via velocity monitoring with commodity {Wi-Fi}},
  author={Qian, Kun and Wu, Chenshu and Yang, Zheng and Liu, Yunhao and Jamieson, Kyle},
  booktitle={Proceedings of the 18th ACM international symposium on mobile ad hoc networking and computing},
  pages={1--10},
  year={2017}
}

@article{Indotrack,
  title={{IndoTrack}: Device-free indoor human tracking with commodity {Wi-Fi}},
  author={Li, Xiang and Zhang, Daqing and Lv, Qin and Xiong, Jie and Li, Shengjie and Zhang, Yue and Mei, Hong},
  journal={Proceedings of the ACM on Interactive, Mobile, Wearable and Ubiquitous Technologies},
  volume={1},
  number={3},
  pages={1--22},
  year={2017},
  publisher={ACM New York, NY, USA}
}

@ARTICLE{Witraj,
  author={Wu, Dan and Zeng, Youwei and Gao, Ruiyang and Li, Shenjie and Li, Yang and Shah, Rahul C. and Lu, Hong and Zhang, Daqing},
  journal={IEEE Transactions on Mobile Computing}, 
  title={{WiTraj}: Robust Indoor Motion Tracking With {WiFi} Signals}, 
  year={2023},
  volume={22},
  number={5},
  pages={3062-3078},
  doi={10.1109/TMC.2021.3133114}}

@ARTICLE{ML_Track,
  author={Shi, Fangzhan and Li, Wenda and Tang, Chong and Fang, Yuan and Brennan, Paul V. and Chetty, Kevin},
  journal={IEEE Transactions on Mobile Computing}, 
  title={{ML-Track}: Passive Human Tracking Using {WiFi Multi-Link Round-Trip CSI} and Particle Filter}, 
  year={2025},
  volume={24},
  number={6},
  pages={5155-5172},
  doi={10.1109/TMC.2025.3529897}}

@article{PassiveSensing,
  title={Awireless passive radar system for real-time through-wall movement detection},
  author={Tan, Bo and Woodbridge, Karl and Chetty, Kevin},
  journal={IEEE Transactions on Aerospace and Electronic Systems},
  volume={52},
  number={5},
  pages={2596--2603},
  year={2016},
  publisher={IEEE}
}

@article{PassiveHand,
  title={Passive handwriting tracking via weak mmWave communication signals},
  author={Yu, Chao and Luo, Yan and Chen, Renqi and Wang, Rui},
  journal={IEEE wireless communications letters},
  volume={13},
  number={3},
  pages={874--878},
  year={2024},
  publisher={IEEE}
}

@ARTICLE{UAV,
  author={Sun, Yifei and Yu, Chao and Luo, Yan and Xiao Han, Tony and Tan, Haisheng and Wang, Rui and Lau, Francis C. M.},
  journal={IEEE Open Journal of the Communications Society}, 
  title={An Experimental Study of Passive {UAV} Tracking With Digital Arrays and Cellular Downlink Signals}, 
  year={2025},
  volume={6},
  number={},
  pages={3779-3794},
  doi={10.1109/OJCOMS.2025.3558430}}

@article{tan2005passive,
  title={Passive radar using global system for mobile communication signal: theory, implementation and measurements},
  author={Tan, Danny KP and Sun, Hongbo and Lu, Yilong and Lesturgie, Marc and Chan, Hian Lim},
  journal={IEE Proceedings-Radar, Sonar and Navigation},
  volume={152},
  number={3},
  pages={116--123},
  year={2005},
  publisher={IET}
}

@article{GN,
author = {Gill, Philip E. and Murray, Walter},
title = {Algorithms for the Solution of the Nonlinear Least-Squares Problem},
journal = {SIAM Journal on Numerical Analysis},
volume = {15},
number = {5},
pages = {977-992},
year = {1978},
doi = {10.1137/0715063},
}

@article{MS_LM,
  title={Trajectory Tracking for MmWave Communication Systems via Cooperative Passive Sensing},
  author={Yu, Chao and Lv, Bojie and Qiu, Haoyu and Wang, Rui},
  journal={ZTE Communications},
  volume={22},
  number={3},
  pages={29–36},
  year={2024},
  doi={10.12142/ZTECOM.202403005}
}

@article{LM-1,
  title={A method for the solution of certain non-linear problems in least squares},
  author={Levenberg, Kenneth},
  journal={Quarterly of applied mathematics},
  volume={2},
  number={2},
  pages={164--168},
  year={1944}
}

@article{LM-2,
  title={An algorithm for least-squares estimation of nonlinear parameters},
  author={Marquardt, Donald W},
  journal={Journal of the society for Industrial and Applied Mathematics},
  volume={11},
  number={2},
  pages={431--441},
  year={1963},
  publisher={SIAM}
}

@inproceedings{EKF_noise,
author = {X. Rong Li and Vesselin P. Jilkov},
title = {{Survey of maneuvering target tracking: dynamic models}},
volume = {4048},
booktitle = {Signal and Data Processing of Small Targets 2000},
editor = {Oliver E. Drummond},
organization = {International Society for Optics and Photonics},
publisher = {SPIE},
pages = {212 -- 235},
year = {2000},
doi = {10.1117/12.391979},
}

@techreport{LM_Hessian,
    author      = {Gavin, Henri P.},
    title       = {The Levenberg–Marquardt algorithm for nonlinear least squares curve-fitting problems},
    institution = {Dept. Civil Environ. Eng., Duke Univ.},
    address     = {Durham, NC, USA},
    month       = {May},
    year        = {2019}
}

@misc{NI_USRP_2954,
    author = "{National Instruments}",
    title = "{USRP - 2954}",
    howpublished = "\url{https://www.ni.com/zh-cn/shop/model/usrp-2954.html}",
    note = "Accessed: 09-May-2025"
}

@misc{Sivers_EvaluationKits,
    author = "{Sivers Semiconductors}",
    title = "{Evaluation Kits (EVK) and Evaluation Boards (EVB)}",
    howpublished = "\url{https://www.sivers-semiconductors.com/5g-millimeter-wave-mmwave-and-satcom/wireless-products/evaluation-kits/}",
    note = "Accessed: 09-May-2025"
}

@misc{zed_camera_webpage,
    title = "{Stereolabs Docs: API Reference, Tutorials, and Integration}",
    howpublished = {[Online]. Available: \url{https://www.stereolabs.com/docs}},
    note = {Accessed: 09-May-2025}
}

@ARTICLE{hybrid_beam,
  author={Molisch, Andreas F. and Ratnam, Vishnu V. and Han, Shengqian and Li, Zheda and Nguyen, Sinh Le Hong and Li, Linsheng and Haneda, Katsuyuki},
  journal={IEEE Communications Magazine}, 
  title={Hybrid Beamforming for Massive MIMO: A Survey}, 
  year={2017},
  volume={55},
  number={9},
  pages={134-141},
  doi={10.1109/MCOM.2017.1600400}}

@inproceedings{multone,
author = {Teng Wei and Anfu Zhou and Xinyu Zhang},
title = {Facilitating Robust 60 {GHz} Network Deployment By Sensing Ambient Reflectors},
booktitle = {14th USENIX Symposium on Networked Systems Design and Implementation (NSDI 17)},
year = {2017},
isbn = {978-1-931971-37-9},
address = {Boston, MA},
pages = {213--226},
publisher = {USENIX Association},
month = mar
}

@ARTICLE{JCIN,
  author={Sun, Yifei and Li, Jie and Zhang, Tong and Wang, Rui and Peng, Xiaohui and Han, Xiao and Tan, Haisheng},
  journal={Journal of Communications and Information Networks}, 
  title={An Indoor Environment Sensing and Localization System via mmWave Phased Array}, 
  year={2022},
  volume={7},
  number={4},
  pages={383-393},
  doi={10.23919/JCIN.2022.10005216}}

@book{MTT,
  title={Design and Analysis of Modern Tracking Systems},
  author={Blackman, Samuel S. and Popoli, Robert J.},
  year={1999},
  publisher={Artech House},
  address={Norwood, MA, USA}
}

\end{document}